\documentclass{elsarticle}
\usepackage[utf8]{inputenc}

\usepackage{sidecap}
\usepackage[algo2e]{algorithm2e}
\usepackage{algorithmic}
\usepackage{algorithm}
\usepackage{float}
\newfloat{algorithm}{t}{lop}
\usepackage{bbold}
\usepackage[bb=ams]{mathalfa}
\usepackage{dsfont}
\usepackage{amsfonts}
\usepackage{bm}
\usepackage{hyperref}
\usepackage{multirow}
\usepackage{setspace}
\usepackage{listings}
\usepackage{pythonhighlight}
\usepackage{graphicx}
\usepackage{amsmath}
\DeclareMathOperator*{\argmax}{arg\,max}

\usepackage{caption}
\usepackage{amsmath}
\usepackage{scalerel}

\makeatletter
\newcommand{\overleftrightsmallarrow}{\mathpalette{\overarrowsmall@\leftrightarrowfill@}}
\newcommand{\overrightsmallarrow}{\mathpalette{\overarrowsmall@\rightarrowfill@}}
\newcommand{\overleftsmallarrow}{\mathpalette{\overarrowsmall@\leftarrowfill@}}
\newcommand{\overarrowsmall@}[3]{%
  \vbox{%
    \ialign{%
      ##\crcr
      #1{\smaller@style{#2}}\crcr
      \noalign{\nointerlineskip}
      $\m@th\hfil#2#3\hfil$\crcr
    }%
  }%
}
\def\smaller@style#1{%
  \ifx#1\displaystyle\scriptstyle\else
    \ifx#1\textstyle\scriptstyle\else
      \scriptscriptstyle
    \fi
  \fi
}
\makeatother
\newcommand{\te}[1]{\overleftrightsmallarrow{#1}}

\newcommand{\CHW}{\ensuremath{C_{\PH\PW}}\xspace}
\newcommand{\CHWtilde}{\ensuremath{C_{\PH\widetilde{\PW}}}\xspace}
\newcommand{\CHQ}{\ensuremath{C_{\PH\textrm{Q}^{(3)}}}\xspace}
\usepackage{definitions}

\journal{arXiv}
\begin{document}

\begin{frontmatter}
\title{Learning the EFT likelihood with tree boosting}

\author[a]{Suman Chatterjee}
\author[b]{Stefan Rohshap}
\author[a]{Robert Sch\"ofbeck}
\author[a]{Dennis Schwarz}

\cortext[x]{\textit{E-mail addresses:} suman.chatterjee@oeaw.ac.at, robert.schoefbeck@oeaw.ac.at,  stefan.rohshap@tuwien.ac.at, dennis.schwarz@oeaw.ac.at}

\address[a]{Institute of High Energy Physics (HEPHY), Austrian Academy of Sciences (\"{O}AW), Nikolsdorfer Gasse 18, 1050 Vienna, Austria}
\address[b]{TU Wien, Karlsplatz 13, 1040 Vienna, Austria}

\date{\today}

\begin{abstract}
We develop a tree boosting algorithm for collider measurements of multiple Wilson coefficients in effective field theories describing phenomena beyond the standard model of particle physics.
The design of the discriminant exploits per-event information of the simulated data sets that encodes the predictions for different values of the Wilson coefficients. 
This ``Boosted Information Tree'' algorithm provides nearly optimal discrimination power order-by-order in the expansion in the Wilson coefficients and approaches the optimal likelihood ratio test statistic.
As a proof-of-principle, we apply the algorithm to the  $\textrm{pp}\rightarrow\PZ\Ph$ process for different types of modeling. 

\end{abstract}


\end{frontmatter}
\clearpage
\tableofcontents
\section{Introduction}\label{sec:Intro}

If the energy scale of physics beyond the standard model (BSM) is higher than the reach of colliders such as the Large Hadron Collider~(LHC), its effects can still be observed in precision measurements of differential distributions as subtle differences to standard model~(SM) predictions.
However, the richness of the large LHC data sets poses a considerable challenge to analysis strategies because both the number of parameters of interest and the number of reconstructed event features affected by BSM phenomena can become very large.

Fortunately, SM effective field theory (SM-EFT)~\cite{Burgess:2007pt,Giudice:2007fh,Grinstein:1991cd,Brivio:2017vri,deFlorian:2016spz,Grzadkowski:2010es}, parameterizing non-resonant deviations below a specific energy scale, mitigates the complexity on the theoretical side. 
Models of SM-EFT~\cite{Degrande:2020evl,Brivio:2020onw} have become the standard choice to describe specific BSM effects in LHC searches. First global interpretations of collider results~\cite{Falkowski:2019hvp,DeBlas:2019ehy,Ellis:2020unq,Dawson:2020oco,Ethier:2021bye,Ethier:2021ydt} already pave the way for establishing SM-EFT as the language of the LHC's legacy.
At mass dimension 6, the SM-EFT has 2499 degrees of freedom (Wilson coefficients), among which 59 are flavor diagonal and respect baryon and lepton number conservation~\cite{Grzadkowski:2010es,Buchmuller:1985jz}.
Constraining this parameter space with the equally high-dimensional feature vectors from the recorded events poses a significant challenge that has led to the development of SM-EFT-specific machine-learning tools. 
Neural networks, in particular, have been used to estimate the likelihood function and related quantities as functions of Wilson coefficients~\cite{Cranmer:2015bka,Brehmer:2018kdj,Brehmer:2018eca,Brehmer:2018hga,Brehmer:2019xox}. Other approaches use neural networks to optimize the variance in parameter regression~\cite{DeCastro:2018psv}, or to learn the cross section ratio to construct an optimal test statistic~\cite{Chen:2020mev}.

Tree boosting algorithms~\cite{Friedman:2001wbq, Friedman:2002we,10.5555/3009657.3009730,Mason99boostingalgorithms}, on the other hand, have received less attention in the context of probing EFT operators, despite robust performance in a wide range of classification problems in high-energy physics.
In a recent work~\cite{Chatterjee:2021nms}, we have introduced the ``Boosted Information Tree''~(BIT), which provides a discriminant for a single-parameter measurement that is statistically optimal as long as the predicted yields vary approximately linearly with the Wilson coefficients.
This linear approximation is valid for numerically small coefficients.
In this work, we extend Boosted Information Trees to the multi-parameter case and extend the applicability to the whole parameter space.

The new developments are based on particle physics event generation, consisting of consecutive steps beginning with the simulation of the hard   SM-EFT interaction, the parton shower, the matrix element~(ME) matching, the hadronization of strongly interacting partons, and the detector response to the stable particles.
Following ideas developed for neural networks in Refs.~\cite{Cranmer:2015bka,Brehmer:2018kdj,Brehmer:2018eca,Brehmer:2018hga,Brehmer:2019xox}, we show in Section~\ref{sec:LL} how the structure of event simulation combined with particular mean squared error~(MSE) loss functionals can be used to regress on the true likelihood using tree boosting, even if the simulated training data is simultaneously conditional on unobservable parton-level ``latent'' features with a non-tractable relation to the detector-level. 
The resulting  ``Boosted Information Tree'' algorithm, discussed in Section~\ref{sec:bit-algo}, allows estimating the differential cross section ratio and the true detector-level likelihood, order by order, in the Wilson-coefficient expansion. 
In many cases, this expansion terminates at a low polynomial order. The Neyman-Pearson lemma then guarantees the discriminator's optimality for simple~(non-composite) hypothesis tests in the entire parameter space.
In Section~\ref{sec:optimality-toy}, this is confirmed using simulation of an analytic model of the $\textrm{pp}\rightarrow\PZ\Ph$ process in a non-trivial but tractable example.
In Section~\ref{sec:binned}, we present a more realistic study of $\PZ\Ph$ production, including the essential backgrounds.

\section{Learning the likelihood from simulation}\label{sec:LL}
\subsection{The optimal test statistic}
Our starting point is the Neyman-Pearson lemma,  whereby the most powerful test statistic to discriminate between two hypotheses $\boldsymbol{\theta}$ and $\boldsymbol{\theta}_0$ is the (negative log-) likelihood ratio
\begin{equation}
q_{\boldsymbol{\theta}}(\mathcal{D})=-\log\frac{L(\mathcal{D}|\boldsymbol{\theta})}{L(\mathcal{D}|\boldsymbol{\theta_0})}.\label{eq:log-likelihood-general}
\end{equation}
The likelihood function $L(\mathcal{D}|\boldsymbol{\theta})$ is the probability to observe a data set $\mathcal{D}=\{\boldsymbol{x}_i\}_{i=1}^N$ of $N$ events, each with a feature vector $\boldsymbol{x}$,  under the hypothesis defined by the vector of model parameters $\boldsymbol{\theta}$. 
In particle physics, it is often possible to write the likelihood function $L(\mathcal{D}|\boldsymbol{\theta})$ as the product of a Poisson contribution, corresponding to the observation of the total number of events $N\sim p(N|\boldsymbol{\theta})=\textrm{P}_{\mathcal{L}\sigma(\boldsymbol{\theta})}(N)$ and a contribution from the normalized probability density function~(pdf), 
\begin{equation}
 p(\boldsymbol{x}|\boldsymbol{\theta})=\frac{1}{\sigma(\boldsymbol{\theta})}\frac{\textrm{d}\sigma_{\boldsymbol\theta}(\boldsymbol{x})}{\textrm{d}\boldsymbol{x}},
\end{equation}
where $\textrm{d}\sigma_{\boldsymbol{\theta}}(\boldsymbol{x})/\textrm{d}\boldsymbol{x}$ is the detector-level differential cross section.
The mean of the Poisson pdf is $\mathcal{L}\sigma(\boldsymbol{\theta})$, where $\sigma(\boldsymbol{\theta})$ is the inclusive cross section and  $\mathcal{L}$ is the integrated luminosity. 
This ``extended'' likelihood function becomes
\begin{equation}
    L(\mathcal{D}|\boldsymbol{\theta})=\textrm{P}_{\mathcal{L}\sigma(\boldsymbol{\theta})}(N)\times\prod_{i=1}^N p(\boldsymbol{x}_i|\boldsymbol{\theta})=\frac{e^{-\mathcal{L}\sigma(\boldsymbol{\theta})}}{N!}\times\prod_{i=1}^N \mathcal{L}\sigma(\boldsymbol{\theta})p(\boldsymbol{x}_i|\boldsymbol{\theta})
\end{equation}
and the optimal statistic follows as
\begin{equation}
q_{\boldsymbol{\theta}}(\mathcal{D})=\mathcal{L}\left(\sigma(\boldsymbol{\theta})-\sigma(\boldsymbol{\theta}_0)\right)-\sum_{i=1}^N\log R(\boldsymbol{x}_i|\boldsymbol{\theta},\boldsymbol{\theta}_0),\label{eq:NLL_teststatistic}
\end{equation}
where
\begin{equation}
R(\boldsymbol{x}|\boldsymbol{\theta},\boldsymbol{\theta}_0)=\frac{\textrm{d}\sigma_{\boldsymbol\theta}(\boldsymbol{x})/\textrm{d}\boldsymbol{x}}{\textrm{d}\sigma_{\boldsymbol\theta_0}(\boldsymbol{x})/\textrm{d}\boldsymbol{x}}=\frac{\sigma(\boldsymbol{\theta})\,p(\boldsymbol{x}|\boldsymbol{\theta})}{\sigma(\boldsymbol{\theta}_0)\,p(\boldsymbol{x}|\boldsymbol{\theta}_0)}
\label{eq:R}
\end{equation}
is the ratio of the differential cross sections computed at the parameter points $\boldsymbol{\theta}$ and $\boldsymbol{\theta}_0$. 
The first term in Eq.~\ref{eq:NLL_teststatistic} is independent of the feature vector $\boldsymbol{x}$ and can be obtained from, e.g., simulation or analytic calculation. 
Because the total cross section $\sigma(\boldsymbol{\theta})$ is available, $R(\boldsymbol{x}|\boldsymbol{\theta},\boldsymbol{\theta}_0)$ equals the detector-level likelihood ratio $r(\boldsymbol{x}|\boldsymbol{\theta},\boldsymbol{\theta}_0)=p(\boldsymbol{x}|\boldsymbol{\theta})/p(\boldsymbol{x}|\boldsymbol{\theta}_0)$ multiplied with a known factor.

We only need to know $R(\boldsymbol{x}|\boldsymbol{\theta},\boldsymbol{\theta}_0)$ to obtain the most powerful test statistic. 
In the following, a tree-boosting regression algorithm, predicting this quantity from the data, is developed. To do so efficiently, we first study how $R(\boldsymbol{x}|\boldsymbol{\theta},\boldsymbol{\theta}_0)$ incurs a simple $\boldsymbol{\theta}$-dependence from the parton-level ME simulation.

\subsection{The training data and the mean-squared error loss functional}
The SM-EFT Lagrangian extends the SM by symmetry-preserving operators $\mathcal{O}_a$ of mass dimension $d_a>4$ to
\begin{equation}
    \mathcal{L}_{\textrm{SM-EFT}}=\mathcal{L}^{(4)}_{\textrm{SM}}+\sum_{a}\frac{\theta_a\mathcal{O}_a}{\Lambda^{d_a-4}}.\label{eq:SMEFT-Lagrangian}
\end{equation}
It is designed to capture non-resonant BSM effects below a UV scale $\Lambda$, conventionally chosen to be $1\TeV$. 
The generic structure of a ME with a single EFT operator insertion and parton-level configuration $\boldsymbol{z}$ is of the general form
\begin{equation}
\textrm{d}\sigma({\boldsymbol\theta})\propto |\mathcal{M}_{\textrm{SM}}({\boldsymbol z})+\theta_a\mathcal{M}^a_{\textrm{BSM}}({\boldsymbol z})|^2 \textrm{d}{\boldsymbol z}.\label{eq:poly-xsec}
\end{equation}
Multiple operator insertions lead to higher-degree polynomials. 
The leading BSM contribution is $2\theta_a\textrm{Re}\left(\mathcal{M}^\ast({\boldsymbol z})_{\textrm{SM}} \mathcal{M}^a_{\textrm{BSM}}({\boldsymbol z})\right)$ and describes the interference of the BSM and SM amplitudes. It is the only term in the expansion in the SM-EFT Wilson coefficients $\boldsymbol{\theta}$ that does not also receive contributions from EFT operators with higher mass dimensions.
With Eq.~\ref{eq:poly-xsec}, we write the parton-level likelihood $p(\boldsymbol{z}|\boldsymbol{\theta})$ as 
\begin{equation}
    p(\boldsymbol{z}|\boldsymbol{\theta})=\frac{1}{\sigma(\boldsymbol{\theta})}\frac{\textrm{d}\sigma_{\boldsymbol{\theta}}( {\boldsymbol z})}{\textrm{d}\boldsymbol{z}},\label{eq:parton-level-likelihood}
\end{equation}
where $\sigma(\boldsymbol{\theta})$ and $\textrm{d}\sigma({\boldsymbol\theta})/\textrm{d}\boldsymbol{z}$ are the total and the differential parton-level cross sections, respectively. Both quantities are polynomial\footnote{Some EFT operators mix with SM kinetic terms and, therefore, can lead to non-polynomial dependence on $\boldsymbol{\theta}$. A polynomial truncation is in many cases possible and offered, e.g., in the \textsc{SMEFTsim} 3.0 model~\cite{Brivio:2020onw}.}  in $\boldsymbol{\theta}$.
Event simulation of the squared MEs at the leading order~(LO) in perturbation theory, in this language, provides an event sample of parton-level configurations ${\boldsymbol z}_i\sim p({\boldsymbol z}|{\boldsymbol \theta}_{\textrm{ref}})$ at some reference parameter point ${\boldsymbol \theta}_{\textrm{ref}}$, that may or may not correspond to the SM at $\boldsymbol{\theta}=0$. 

It is important to realize that the per-event ME in Eq.~\ref{eq:poly-xsec} can easily be reevaluated for different $\boldsymbol{\theta}$ once the $\boldsymbol{\theta}$-independent ME terms $\mathcal{M}_{\textrm{SM}}(\boldsymbol{z})$ and $\mathcal{M}_{\textrm{BSM}}^a(\boldsymbol{z})$ are known for a specific $\boldsymbol{z}$.
Because the differential cross section is polynomial,  a sufficient number of linearly independent $\boldsymbol{\theta}$ is enough to infer the full $\boldsymbol{\theta}$-dependence for every simulated $\boldsymbol{z}$ from the event generator. 
Therefore, we can obtain analytically known per-event weight functions $w_i(\boldsymbol{\theta})$~\cite{Brehmer:2018eca} that encode the $\boldsymbol{\theta}$-dependence of all simulated predictions.
For $N_{\textrm{c}}$ different coefficients and $N_{\textrm{i}}$ operator insertions, a minimum of
\begin{equation}\frac{2N_{\textrm{i}}+1}{N_{\textrm{c}}}\binom{2N_{\textrm{i}}+N_{\textrm{c}}}{2N_{\textrm{i}}+1}\end{equation}
of such evaluations are necessary. The polynomial degree, in general, is $2N_{\textrm{i}}$.
Choosing an overall normalization~(scaling) to the total number of predicted events for an arbitrary luminosity $\mathcal{L}$, we get the approximation
\begin{equation}
    \int_{\Delta \boldsymbol{z}}\frac{\textrm{d}\sigma_ {\boldsymbol\theta}(\boldsymbol{z})}{\textrm{d}\boldsymbol{z}}\textrm{d}\boldsymbol{z}\approx\frac{\textrm{d}\sigma_ {\boldsymbol\theta}(\boldsymbol{z})}{\textrm{d}\boldsymbol{z}}\Delta \boldsymbol{z}\approx\frac{1}{\mathcal{L}}\sum_{z_i\in\Delta \boldsymbol{z}} w_i(\boldsymbol{\theta})\label{eq:diff-xsec-approx}
\end{equation} 
for the differential parton-level cross section, integrated over a small phase space volume $\Delta \boldsymbol{z}$.
Integrating over $\boldsymbol{z}$ leads to the relation
\begin{equation}
    \sum_{i=1}^{N_{\textrm{sim}}} w_i(\boldsymbol{\theta})=\mathcal{L}\sigma(\boldsymbol{\theta})
\end{equation}
for the total cross section.
The approximation in Eq.~\ref{eq:diff-xsec-approx} is valid for a sufficiently large simulated data set and when $\textrm{d}\sigma_ {\boldsymbol\theta}(\boldsymbol{z})/\textrm{d}\boldsymbol{z}$ does not vary strongly in the region $\Delta \boldsymbol{z}$.

At the LO, we can furthermore relate the weight functions with the underlying probabilistic model as 
\begin{equation}
    w_i(\boldsymbol{\theta})=\mathcal{L}\,\frac{\textrm{d}\sigma_{\boldsymbol{\theta}}(\boldsymbol{z})}{\textrm{d}\boldsymbol{z}}\Big|_{\boldsymbol{z}=\boldsymbol{z}_i}=\mathcal{L}\sigma(\boldsymbol{\theta})\,p(\boldsymbol{z}_i|\boldsymbol{\theta})\label{eq:wi-definition}.
\end{equation}
In contrast to the ME, the simulation of the parton shower, the ME matching, the hadronization of strongly interacting partons, and the detector response to the stable particles are all intractable.
These steps provide each event with an (in principle) observable simulated detector-level feature vector ${\boldsymbol x}$ such that each pair $(\boldsymbol{x}_i,\boldsymbol{z}_i)$ in the simulated data set  $\mathcal{D}_{\textrm{sim}} = \{\boldsymbol{x}_i,\boldsymbol{z}_i,w_i(\boldsymbol{\theta})\}_{i=1}^{N_\textrm{sim}}$ is drawn from their joint probability distribution,  $(\boldsymbol{x}_i,\boldsymbol{z}_i)\sim p(\boldsymbol{x},\boldsymbol{z}|\boldsymbol{\theta}_{\textrm{ref}})$. 
For a detector-level bin $\Delta\boldsymbol{x}$, we then obtain the expected Poisson yield from the simulated sample as $\lambda(\boldsymbol{\theta})=\sum_{\boldsymbol{x}_i\in \Delta\boldsymbol{x}}w_i(\boldsymbol\theta)$. 

The detector-level likelihood, appearing in the Neyman-Pearson lemma, is related to the joint likelihood by the factorization
\begin{equation}
p(\boldsymbol{x},\boldsymbol{z}|\boldsymbol{\theta})=p(\boldsymbol{x}|\boldsymbol{z})p(\boldsymbol{z}|\boldsymbol{\theta}),\label{eq:joint-L-factorization}
\end{equation}
which holds if the probability $p(\boldsymbol{x}|\boldsymbol{z})$ to observe a feature vector $\boldsymbol{x}$ given a parton-level configuration $\boldsymbol{z}$ is independent of the EFT parameters~$\boldsymbol{\theta}$, including possible loop corrections.  
This assumption is valid in many cases but excludes, for example, modifications of the strong interaction entering the modelling of the parton shower. 
The configuration spaces of both, observed event features $\boldsymbol{x}$ and parton-level configuration $\boldsymbol{z}$ are high-dimensional with large numbers of continuous and discrete variables.  
Equation~\ref{eq:joint-L-factorization} implies the cancellation of the intractable factor in the joint likelihood-ratio
\begin{equation}
    r(\boldsymbol{x},\boldsymbol{z}|\boldsymbol{\theta},\boldsymbol{\theta}_0)\equiv\frac{p(\boldsymbol{x},\boldsymbol{z}|\boldsymbol{\theta})}{p(\boldsymbol{x},\boldsymbol{z}|\boldsymbol{\theta}_0)}=\frac{p(\boldsymbol{z}|\boldsymbol{\theta})}{p(\boldsymbol{z}|\boldsymbol{\theta}_0)}
\end{equation}
so that we have the LO relation
\begin{eqnarray}
r(\boldsymbol{x}_i,\boldsymbol{z}_i|\boldsymbol{\theta},\boldsymbol{\theta}_0)=\frac{\sigma(\boldsymbol{\theta}_0)}{\sigma(\boldsymbol{\theta})}\frac{w_i(\boldsymbol{\theta})}{w_i(\boldsymbol{\theta}_0)}\label{eq:r-in-terms-of-wi}
\end{eqnarray}
for any simulated event. 
The situation is different when we try to evaluate the likelihood at the detector level $p(\boldsymbol{x}|\boldsymbol{\theta})$,  appearing in the cross section ratio $R(\boldsymbol{x}|\boldsymbol{\theta},\boldsymbol{\theta}_0)$ in Eq.~\ref{eq:R}.
We need to estimate the detector-level likelihood
\begin{eqnarray}
    p(\boldsymbol{x}|\boldsymbol{\theta})=\int\,p(\boldsymbol{x},\boldsymbol{z})\textrm{d}\boldsymbol{z}=\int\,p(\boldsymbol{x}|\boldsymbol{z})p(\boldsymbol{z}|\boldsymbol{\theta})\textrm{d}\boldsymbol{z}\label{eq:factorizedLL},
\end{eqnarray}
which contains an integral over the intractable $p(\boldsymbol{x}|\boldsymbol{z})$  that neither cancels in a detector-level likelihood ratio nor can it be easily reevaluated for different $\boldsymbol{z}$ at the same $\boldsymbol{x}$. 

How can we use all this structure to construct an estimator of $R(\boldsymbol{x}|\boldsymbol{\theta},\boldsymbol{\theta}_0)$ given only joint quantities, also conditional on the parton-level $\boldsymbol{z}$, and including the integration over $p(\boldsymbol{x}|\boldsymbol{z})$? 
The key point from Refs.~\cite{Brehmer:2018eca,Cranmer:2015bka,Brehmer:2018kdj,Brehmer:2018hga,Brehmer:2019xox} is to use certain mean-squared error~(MSE) loss functions that can be shown to regress on functions of $\boldsymbol{x}$, for example $p(\boldsymbol{x}|\boldsymbol{\theta})$ or $r(\boldsymbol{x}|\boldsymbol{\theta},\boldsymbol{\theta}_0)$, given only  joint quantities.
In the main line of reasoning, an MSE loss functional
\begin{eqnarray}
\textrm{MSE}[\hat G]=\int \textrm{d}\boldsymbol{x}\,\textrm{d}\boldsymbol{z}\, p(\boldsymbol{x},\boldsymbol{z}|\boldsymbol{\theta}_0)\left(F(\boldsymbol{x},\boldsymbol{z})-\hat F(\boldsymbol{x})\right)^2\label{eq:MSE-formally}
\end{eqnarray}
with the joint pdf $p(\boldsymbol{x},\boldsymbol{z}|\boldsymbol{\theta}_0)$ is used to fit an estimator $\hat F(\boldsymbol{x})$ to some quantity of choice $F(\boldsymbol{x},\boldsymbol{z})$.
While the estimator $\hat F(\boldsymbol{x})$ only depends on the detector-level observables $\boldsymbol{x}$, the target of the regression can also depend on $\boldsymbol{z}$. The loss functional
is formally minimized by
\begin{eqnarray}
F^\ast(\boldsymbol{x})=\frac{\int \textrm{d}\boldsymbol{z}\, p(\boldsymbol{x},\boldsymbol{z}|\boldsymbol{\theta}_0)F(\boldsymbol{x},\boldsymbol{z})}{\int \textrm{d}\boldsymbol{z}\, p(\boldsymbol{x},\boldsymbol{z}|\boldsymbol{\theta}_0)}.\label{eq:g-ast-formally}
\end{eqnarray}
Choosing, for example, $F(\boldsymbol{x},\boldsymbol{z})=r(\boldsymbol{x},\boldsymbol{z}|\boldsymbol{\theta},\boldsymbol{\theta}_0)$ we obtain
\begin{eqnarray}
F^\ast(\boldsymbol{x})=\frac{\int \textrm{d}\boldsymbol{z}\, p(\boldsymbol{x},\boldsymbol{z}|\boldsymbol{\theta})}{\int \textrm{d}\boldsymbol{z}\, p(\boldsymbol{x},\boldsymbol{z}|\boldsymbol{\theta}_0)}=\frac{p(\boldsymbol{x}|\boldsymbol{\theta})}{p(\boldsymbol{x}|\boldsymbol{\theta}_0)}=r(\boldsymbol{x}|\boldsymbol{\theta},\boldsymbol{\theta}_0),
\end{eqnarray}
i.e., the loss functional in Eq.~\ref{eq:MSE-formally}, calculable from Eq.~\ref{eq:r-in-terms-of-wi}, provides an $F^\ast(\boldsymbol{x})$ regressing on the true detector-level likelihood ratio. 

\subsection{The formal solution in the \texorpdfstring{$\theta$}{theta}-expansion}

Next, we exploit that Eq.~\ref{eq:poly-xsec} implies the finite-order polynomial  of the detector-level likelihood ratio 
\begin{eqnarray}
R(\boldsymbol{x}|\boldsymbol{\theta},\boldsymbol{\theta}_0)&=&1+(\theta-\theta_0)_a\left.\frac{\partial_a(\sigma_{\boldsymbol{\theta}}p(\boldsymbol{x}|\boldsymbol{\theta}))}{\sigma_{\boldsymbol{\theta}}p(\boldsymbol{x}|\boldsymbol{\theta})}\right|_{\boldsymbol{\theta}=\boldsymbol{\theta}_0}\nonumber\\
&&+\,\frac{1}{2}(\theta-\theta_0)_a(\theta-\theta_0)_b\left.\frac{\partial_a\partial_b(\sigma_{\boldsymbol{\theta}}p(\boldsymbol{x}|\boldsymbol{\theta}))}{\sigma_{\boldsymbol{\theta}}p(\boldsymbol{x}|\boldsymbol{\theta}))}\right|_{\boldsymbol{\theta}\equiv\boldsymbol{\theta}_0}\nonumber\\
&=&1+(\theta-\theta_0)_a R_a(\boldsymbol{x}) +\frac{1}{2}(\theta-\theta_0)_a(\theta-\theta_0)_b R_{ab}(\boldsymbol{x})\label{eq:R-poly-expansion}
\end{eqnarray}
for single operator insertions. 
If there are $N_{\textrm{c}}$ components in $\boldsymbol{\theta}$, we have $N_{\textrm{c}}$ functions $R_a(\boldsymbol{x})$ and $N_{\textrm{c}}(N_{\textrm{c}}+1)/2$ different functions $R_{ab}(\boldsymbol{x})$. 

We can formally estimate the functions $R_a(\boldsymbol{x})$ and $R_{ab}(\boldsymbol{x})$ by specifying $F(\boldsymbol{x},\boldsymbol{z})$ to be any of the coefficient functions
\begin{align}
F_a(\boldsymbol{x},\boldsymbol{z})&=\frac{\partial_a\left(\sigma(\boldsymbol{\theta})\,p(\boldsymbol{x},\boldsymbol{z}|\boldsymbol{\theta})\right)\Big|_{\boldsymbol{\theta}=\boldsymbol{\theta}_0}}{\sigma(\boldsymbol{\theta}_0)\,p(\boldsymbol{x},\boldsymbol{z}|\boldsymbol{\theta}_0)\phantom{\Big|_{\boldsymbol{\theta}=\boldsymbol{\theta}_0}}},\label{eq:F-a-xz}\\
F_{ab}(\boldsymbol{x},\boldsymbol{z})&=\frac{\partial_a\partial_b\left(\sigma(\boldsymbol{\theta})\,p(\boldsymbol{x},\boldsymbol{z}|\boldsymbol{\theta})\right)\Big|_{\boldsymbol{\theta}=\boldsymbol{\theta}_0}}{\sigma(\boldsymbol{\theta}_0)\,p(\boldsymbol{x},\boldsymbol{z}|\boldsymbol{\theta}_0)\phantom{\Big|_{\boldsymbol{\theta}=\boldsymbol{\theta}_0}}},\label{eq:F-ab-xz}
\end{align}
jointly depending on $(\boldsymbol{x},\boldsymbol{z})$.
Following again the functional minimization in Ref.~\cite{Brehmer:2018eca}, we find
\begin{align}
F^\ast_a(\boldsymbol{x})&=\frac{\int\textrm{d}\boldsymbol{z}\,\partial_a(\sigma(\boldsymbol{\theta})\,p(\boldsymbol{x},\boldsymbol{z}|\boldsymbol{\theta}))\Big|_{\boldsymbol{\theta}=\boldsymbol{\theta}_0}}{\int\textrm{d}\boldsymbol{z}\,\sigma(\boldsymbol{\theta}_0)\,p(\boldsymbol{x},\boldsymbol{z}|\boldsymbol{\theta}_0)\phantom{\Big|_{\boldsymbol{\theta}=\boldsymbol{\theta}_0}}}=\frac{\partial_a(\sigma(\boldsymbol{\theta})\, p(\boldsymbol{x}|\boldsymbol{\theta}))\Big|_{\boldsymbol{\theta}=\boldsymbol{\theta}_0}}{\sigma(\boldsymbol{\theta}_0)\,p(\boldsymbol{x}|\boldsymbol{\theta}_0)\phantom{\Big|_{\boldsymbol{\theta}=\boldsymbol{\theta}_0}}},\label{eq:Fa-star}\\
F^\ast_{ab}(\boldsymbol{x})&=\frac{\int\textrm{d}\boldsymbol{z}\,\partial_a\partial_b(\sigma(\boldsymbol{\theta})\,p(\boldsymbol{x},\boldsymbol{z}|\boldsymbol{\theta}))\Big|_{\boldsymbol{\theta}=\boldsymbol{\theta}_0}}{\int\textrm{d}\boldsymbol{z}\,\sigma(\boldsymbol{\theta}_0)\,p(\boldsymbol{x},\boldsymbol{z}|\boldsymbol{\theta}_0)\phantom{\Big|_{\boldsymbol{\theta}=\boldsymbol{\theta}_0}}}=\frac{\partial_a\partial_b(\sigma(\boldsymbol{\theta})\, p(\boldsymbol{x}|\boldsymbol{\theta}))\Big|_{\boldsymbol{\theta}=\boldsymbol{\theta}_0}}{\sigma(\boldsymbol{\theta}_0)\,p(\boldsymbol{x}|\boldsymbol{\theta}_0)\phantom{\Big|_{\boldsymbol{\theta}=\boldsymbol{\theta}_0}}},\label{eq:Fab-star}
\end{align}
which are just the coefficients of a Taylor expansion of Eq.~\ref{eq:R} around $\boldsymbol{\theta}_0$.
We can identify $F_a^\ast(\boldsymbol{x})=R_a(\boldsymbol{x})$ and  $F_{ab}^\ast(\boldsymbol{x})=R_{ab}(\boldsymbol{x})$. 
For a higher number of operator insertions $N_\textrm{i}$, the degree of the polynomial expansion in Eq.~\ref{eq:R-poly-expansion} is $2N_\textrm{i}$. 
Because there is no partial integration involved, the formulae extend straightforwardly to these higher-order terms by replacing $\partial_{a}\partial_{b}\rightarrow\partial_{a}\partial_{b}\partial_{c}$, etc.

So far, we have shown that a formal minimization of the loss function in Eq.~\ref{eq:MSE-formally} with $F(\boldsymbol{x},\boldsymbol{z})=F_{a}(\boldsymbol{x},\boldsymbol{z})$ produces the coefficients $R_{a}(\boldsymbol{x})$ in the expansion in Eq.~\ref{eq:R-poly-expansion} and analogously for the coefficient pairs $ab$, etc.
If Eq.~\ref{eq:R-poly-expansion} terminates or can be truncated to good accuracy, we have formally obtained the optimal Neyman-Pearson test statistic of the extended likelihood for arbitrary hypothesis tests of $\boldsymbol{\theta}$ and $\boldsymbol{\theta}_0$. 
Next, we proceed to derive the boosting algorithm, learning the test statistic from simulation. 

\section{Learning the likelihood with tree boosting}\label{sec:bit-algo}

We begin with evaluating Eqs.~\ref{eq:F-a-xz}--\ref{eq:F-ab-xz} for a simulated data set as
\begin{align}
F_{a}(\boldsymbol{x}_i,\boldsymbol{z}_i)&=\frac{\partial_{a}w_i(\boldsymbol{\theta})\Big|_{\boldsymbol{\theta}=\boldsymbol{\theta_0}}}{w_i(\boldsymbol{\theta}_0)\phantom{\Big|_{\boldsymbol{\theta}=\boldsymbol{\theta_0}}}}=\frac{w_{i,a}}{w_{i,0}},\label{eq:Fa-joint}\\
F_{ab}(\boldsymbol{x}_i,\boldsymbol{z}_i)&=\frac{\partial_{a}\partial_b w_i(\boldsymbol{\theta})\Big|_{\boldsymbol{\theta}=\boldsymbol{\theta_0}}}{w_i(\boldsymbol{\theta}_0)\phantom{\Big|_{\boldsymbol{\theta}=\boldsymbol{\theta_0}}}}=\frac{w_{i,ab}}{w_{i,0}},\label{eq:Fab-joint}
\end{align}
where we also Taylor-expand the polynomial per-event weight functions as
\begin{equation}
w_i(\boldsymbol{\theta})=w_{i,0}+(\theta-\theta_0)_a w_{i,a} +\frac{1}{2}(\theta-\theta_0)_a(\theta-\theta_0)_b w_{i,ab},\label{eq:wi-poly-expansion}
\end{equation}
with coefficients $w_{i,0}$, $w_{i,a}$, and $w_{i,ab}$. 
These constants are numerically known for each simulated event. 
For a simulated dataset $\mathcal{D}_{\textrm{sim}}=\{\boldsymbol{x}_i,\boldsymbol{z}_i,w_i(\boldsymbol{\theta})\}_{i=1}^{N_{\textrm{sim}}}$, the MSE loss functions are
\begin{align}
\textrm{MSE}[\hat F_{a}]&\;=\sum_{(\boldsymbol{x},\boldsymbol{z},w)_i\in\mathcal{D}} w_{i,0}\left|\frac{w_{i,a}}{w_{i,0}} - \hat F_{a}(\boldsymbol{x}_i)\right|^2,\label{eq:MSE-linear}\\
\textrm{MSE}[\hat F_{ab}]&\;=\sum_{(\boldsymbol{x},\boldsymbol{z},w)_i\in\mathcal{D}} w_{i,0}\left|\frac{w_{i,ab}}{w_{i,0}} - \hat F_{ab}(\boldsymbol{x}_i)\right|^2.\label{eq:MSE-quad}
\end{align}
These loss functionals are independently minimized for each of the $N_{\textrm{c}}$ linear coefficients labeled by $a$ and the $N_{\textrm{c}}(N_{\textrm{c}}+1)/2$ coefficient pairs labeled by $ab$. 

\subsection{The weak learner}

Boosting, in general, provides a strong learner by iteratively training an ensemble of weak learners to the pseudo-residuals of the previous iteration step.
In our case, the weak learners are regression trees. 
Each is defined by a maximum number $D$ of consecutive requirements on the vector of input features $\boldsymbol{x}$, collectively denoted by $\alpha_j$, that group the input data in no more than $2^D$ terminal nodes $j$, collectively denoted by $\mathcal{J}$. 
All terminal nodes are disjoint and the union of all $j\in\mathcal{J}$ equals the total feature space. 
Each terminal node is associated with a number $F_j$ corresponding to the prediction for an event in $j$. 
The training minimizes the loss function with respect to $\alpha_j$ and $F_j$.
The prediction of a generic tree can be written as
\begin{equation}
\hat F(\mathbf{x})=\sum_{j\in \mathcal{J}}\mathds{1}_{\alpha_j}(\mathbf{x})F_j,
\end{equation}
where $\mathds{1}_{\alpha_j}(\mathbf{x})=1$ if the feature vector $\mathbf{x}$ satisfies the requirements $\alpha_j$ of the terminal node $j$ and is zero otherwise. 
If we use this ansatz for, e.g., the linear terms labeled by $a$ in Eq.~\ref{eq:MSE-linear}, we obtain
\begin{align}
   \textrm{MSE}[\hat F_{a}] &=\sum_{j\in\mathcal{J}}\sum_{i\in j}w_i\left|\frac{w_{i,a}}{w_i}-F_j\right|^2\nonumber\\
&=\sum_{i=1}^{N_{\textrm{sim}}}\frac{w_{i,a}^2}{w_i}-2\sum_{j\in\mathcal{J}}F_j\sum_{i\in j}w_{i,a}+\sum_{j\in\mathcal{J}}F_j^2\sum_{i\in j}w_i.\label{eq:MSE-before-replacing-F}
\end{align}
We can drop the first term because it does not depend on the weak learner's configuration $\mathcal{J}$.
Moreover, we can minimize $ \textrm{MSE}[\hat F_{a}]$ with respect to the $F_j$ and find
\begin{equation}
    F_j=\frac{\sum_{i\in j}w_{i,a}}{\sum_{i\in j}w_i}=\left.\frac{\partial_a\lambda_j}{\lambda_j}\right|_{\boldsymbol{\theta}=\boldsymbol{\theta}_0}=\partial_a\log\lambda_j\bigg|_{\boldsymbol{\theta}=\boldsymbol{\theta}_0}\label{eq:Fj-lin}
\end{equation}
which then leads to
\begin{equation}
   \textrm{MSE}[\hat F_{a}] =-\sum_{j\in\mathcal{J}}\frac{\left(\sum_{i\in j}w_{i,a}\right)^2}{\sum_{i\in j}w_i}=-\sum_{j\in\mathcal{J}}\frac{\left(\partial_a\lambda_j\right)^2}{\lambda_j}\Bigg|_{\boldsymbol{\theta}=\boldsymbol{\theta}_0}=-\sum_{j\in\mathcal{J}}I^{(\lambda_j)}.\label{eq:FI-loss}\\
\end{equation}
As before, $\lambda_j$ labels the simulated Poisson mean in the node $j$ and $I^{(\lambda_j)}$ denotes the Fisher information of a measurement of $\theta_a$ in a Poisson counting experiment with yield $\lambda_j$.
This last equation establishes the correspondence with our earlier work in Ref.~\cite{Chatterjee:2021nms}, where the optimization of the Fisher information was the starting point. 
Because we could eliminate the regression predictions $F_j$, the computational complexity reduces from a regression problem to a classification problem. 

The remaining algorithmic goal for the weak learner is to find the feature-space partitioning $\mathcal{J}$ that minimizes the loss function in Eq.~\ref{eq:FI-loss}. 
The resulting procedure is an adaption of the well-known ``Classification And Regression Tree'' (CART) algorithm~\cite{breiman1984classification}, and we describe it in detail in Ref.~\cite{Chatterjee:2021nms}, including a discussion of computational complexity and overtraining. 
For completeness and reflecting minimal notational changes for the higher-order terms, we include it in \mbox{\ref{app:weak-lerner}}. 

The derivation for the higher-order terms is exactly analogous.
It is enough to replace $w_{i,a}\rightarrow w_{i,ab}$ in Eq.~\ref{eq:MSE-before-replacing-F} and performing the elimination of the regression predictions according to
\begin{equation}
    F_j=\frac{\sum_{i\in j}w_{i,ab}}{\sum_{i\in j}w_i}=\left.\frac{\partial_a\partial_b\lambda_j}{\lambda_j}\right|_{\boldsymbol{\theta}=\boldsymbol{\theta}_0}.\label{eq:Fj-quad}
\end{equation}
The result for the weak learner's loss function is 
\begin{eqnarray}
   \textrm{MSE}[\hat F_{ab}] =-\sum_{j\in\mathcal{J}}\frac{\left(\sum_{i\in j}w_{i,ab}\right)^2}{\sum_{i\in j}w_i}=-\sum_{j\in\mathcal{J}}\frac{\left(\partial_a\partial_b\lambda_j\right)^2}{\lambda_j}\Bigg|_{\boldsymbol{\theta}=\boldsymbol{\theta}_0},\label{eq:MSE-loss-quadratic}
\end{eqnarray}
where the second term implies that the weak learner for the second-order polynomial coefficient function is the same as before, except that it is fed with $w_{i,ab}$ instead of $w_{i,a}$. 
There is no analog to the Fisher information in this case. 

Furthermore, we learn from Eq.~\ref{eq:Fj-lin} and Eq.~\ref{eq:Fj-quad} that our choices in Eqs.~\ref{eq:Fa-joint}--\ref{eq:Fab-joint} and Eqs.~\ref{eq:MSE-linear}--\ref{eq:MSE-quad} indeed lead to an approximation of $R_a(\boldsymbol{x})$ and $R_{ab}(\boldsymbol{x})$ because a direct calculation shows that 
\begin{eqnarray}
    \frac{\mathds{E}_{\boldsymbol{\theta}_0}\left(\mathds{1}_j(\boldsymbol{x})\cdot R_a(\boldsymbol{x})\right)}{\mathds{E}_{\boldsymbol{\theta_0}}\left(\mathds{1}_j(\boldsymbol{x})\right)} &=& \left.\frac{\partial_a\lambda_j}{\lambda_j}\right|_{\boldsymbol{\theta}=\boldsymbol{\theta}_0}\;\;\textrm{and}\\
    \frac{\mathds{E}_{\boldsymbol{\theta}_0}\left(\mathds{1}_j(\boldsymbol{x})\cdot R_{ab}(\boldsymbol{x})\right)}{\mathds{E}_{\boldsymbol{\theta}_0}\left(\mathds{1}_j(\boldsymbol{x})\right)} &=& \left.\frac{\partial_a\partial_b\lambda_j}{\lambda_j}\right|_{\boldsymbol{\theta}=\boldsymbol{\theta}_0},
\end{eqnarray}
which agree with the formal regression prediction in Eq.~\ref{eq:Fj-lin} and Eq.~\ref{eq:Fj-quad}.

\begin{algorithm}[t]
\KwData{Data set $\mathcal{D} = \{\mathbf{x}_i,w_{i,0},w_{a,i}\}_{i=1}^N$ for fixed $a$ or $\mathcal{D} = \{\mathbf{x}_i,w_{i,0},w_{ab,i}\}_{i=1}^N$ for fixed $ab$.}
\KwIn{Number of boosting iterations $B$, learning rate $\eta$}
 \KwOut{Boosted learner $F^{(B)}$ \vspace{0.02cm}}
$F^{(0)} \gets 0$
 \algorithmiccomment{initialize boosted learner}\;
\For{$b \gets 1,\hdots,B$}{
    $f^{(b)} \gets \mathsf{fit}\left(\left\{ \mathbf{x}_i, w_{i,0}, w_{a(,b),i} -\eta w_{i,0}F^{(b-1)}(\mathbf{x}_i) \right\}_{i=1}^{{N}}\right)$ \;
     $F^{(b)} \gets F^{(b-1)}+\eta f^{(b)}$\;
}
\caption{Boosted Information Tree up to quadratic order.}
\label{algo:bit}
\end{algorithm}

\subsection{The boosting algorithm}

The last step is the boosting algorithm, providing a strong learner by iteratively training an ensemble of weak learners to the pseudo-residuals of the previous iteration step. 
We choose a learning rate $\eta$, amounting to the fraction of the prediction we retain from the previous boosting iteration. Values in the range of 10\%--30\% have proven practical.
Assuming that the tree has been trained up to $\hat F^{(b-1)}(\boldsymbol{x})$, we choose $\hat F^{(b)}(\boldsymbol{x})= \hat f^{(b)}(\boldsymbol{x})+\eta \hat F^{(b-1)}(\boldsymbol{x})$ in the weak learner's loss functions in Eqs.~\ref{eq:MSE-linear}--\ref{eq:MSE-quad} for iteration $b$.
We minimize the loss with the configuration of $\hat f^{(b)}(\boldsymbol{x})$, keeping $\hat F^{(b-1)}(\boldsymbol{x})$ fixed. 
The loss function at boosting iteration $b$ is, thus,
\begin{eqnarray}
\textrm{MSE}[\hat f_{a}^{(b)}]&=&\sum_{(\boldsymbol{x},\boldsymbol{z},w)_i\in\mathcal{D}} w_{i,0}\left|\frac{w_{i,a}}{w_{i,0}} - \eta \hat F_{a}^{(b-1)}(\boldsymbol{x}_i)- \hat f_{a}^{(b)}(\boldsymbol{x}_i)\right|^2\nonumber\\
&=&\sum_{(\boldsymbol{x},\boldsymbol{z},w)_i\in\mathcal{D}} w_{i,0}\left|\frac{w_{i,a}- \eta w_{i,0}\hat F_{a}^{(b-1)}(\boldsymbol{x}_i)}{w_{i,0}}- \hat f_{a}^{(b)}(\boldsymbol{x}_i)\right|^2.\label{eq:MSE-boosting}
\end{eqnarray}
The iteration starts at $b=1$ with $\hat F^{(0)}(\boldsymbol{x})=0$ and it stops when $b$ reaches a number $B$, chosen beforehand. 
At $b=1$, the loss function of the weak learner in Eq.~\ref{eq:MSE-linear} is reproduced. 
For $b>1$, the weak learner is fit to $w_{i,a}\rightarrow w_{i,a} -\eta w_{i,0}\hat F^{(b-1)}(\boldsymbol{x}_i)$ and this replacement implements the change of the weight derivatives for the next boosting iteration.

The trivial replacement $w_{i,a}\rightarrow w_{i,ab}$ provides the boosting loss for the higher-order terms. 
The prediction for the cross section ratio at phase space point $\boldsymbol{x}$ is then given by 
\begin{equation}
\hat R(\boldsymbol{x}|\boldsymbol{\theta},\boldsymbol{\theta}_0)=1+(\theta-\theta_0)_a \hat F_a^{(B)}(\boldsymbol{x}) +\frac{1}{2}(\theta-\theta_0)_a(\theta-\theta_0)_b \hat F^{(B)}_{ab}(\boldsymbol{x}),\label{eq:F-poly-estimation}
\end{equation}
which is again a quadratic polynomial in $\boldsymbol{\theta}$.

This completes the construction of the algorithm. 
Because the training is independent for each Wilson coefficient $a$ or pair of Wilson coefficients $ab$, the total number of boosting iterations $B$ can be chosen differently in each case. 
From now on, we drop it from the notation if it does not make a difference. 
We summarise the procedures in Algorithm~\ref{algo:bit}.
The BIT test statistic $\hat q$ is obtained from Eq.~\ref{eq:NLL_teststatistic} by replacing the true cross section ratio $R$ with the estimate $\hat R$,
\begin{equation}
\hat q_{\boldsymbol{\theta}}(\mathcal{D})=\mathcal{L}\left(\sigma(\boldsymbol{\theta})-\sigma(\boldsymbol{\theta}_0)\right)-\sum_{i=1}^N\log \hat R(\boldsymbol{x}_i|\boldsymbol{\theta},\boldsymbol{\theta}_0).\label{eq:BIT-statistic}
\end{equation}

Finally, we note that NLO event simulation can contain negative weights~\cite{Frixione:2002ik} and a rigorous interpretation of a (weighted) event sampling is not necessarily possible. 
While the $w_i(\boldsymbol{\theta})$ can be obtained in a straightforward way, the interpretation in Eqs.~\ref{eq:wi-definition} and \ref{eq:r-in-terms-of-wi} is not generally valid. 
Physically meaningful differential cross sections, however, are always positive and, therefore, the Monte-Carlo approximation in Eq.~\ref{eq:diff-xsec-approx} holds, providing a polynomial that is positive for all $\boldsymbol{\theta}$. 
That is sufficient for Eq.~\ref{eq:FI-loss} and Eq.~\ref{eq:MSE-loss-quadratic} to hold, as long as the minimum terminal node size is chosen large enough to ensure a positive yield in each node. We defer the study of NLO event simulation to future work.

\section{Optimality in toy data}\label{sec:optimality-toy}

\begin{figure}\centering
        \includegraphics[width=0.65\textwidth]{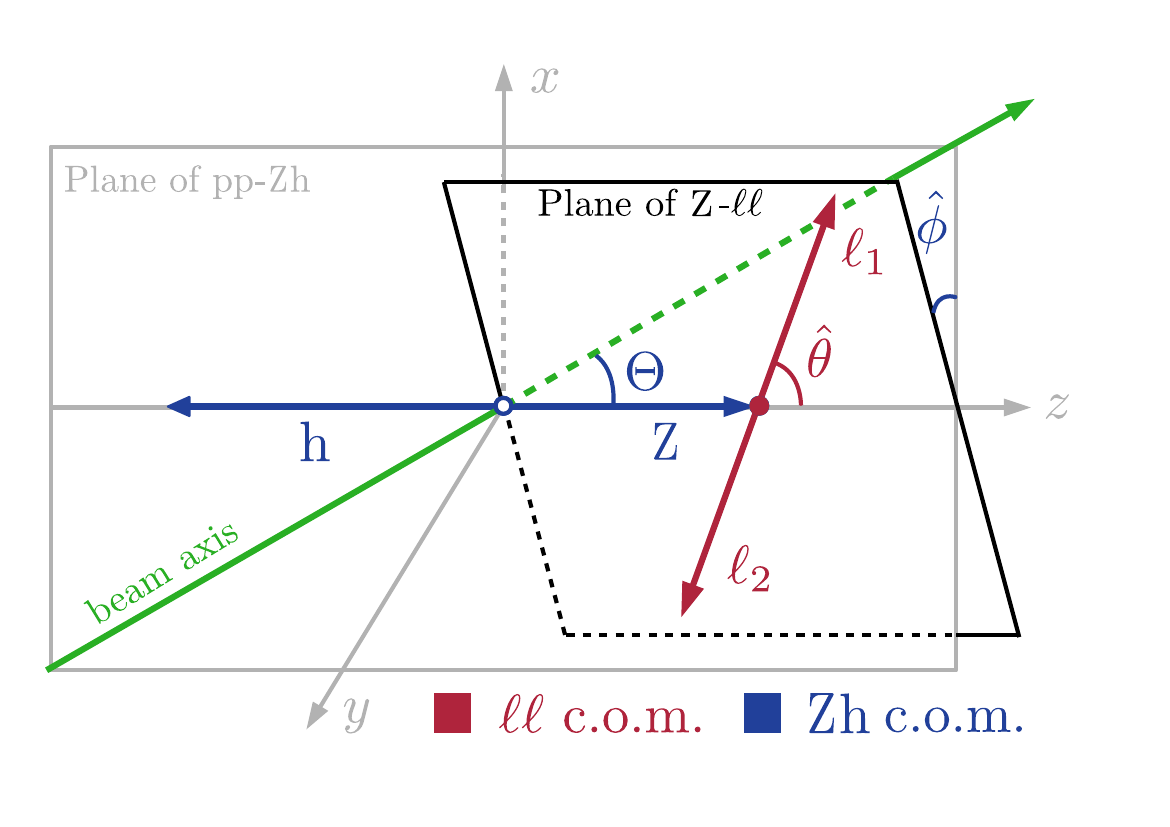}
    \caption{Sketch of the decay planes of the $\PZ\Ph$ process. The angle $\Theta$ between the pp beam axis and the $\PZ\textrm{h}$ system is measured in the $\PZ\Ph$ c.o.m. system, while the the angles $\hat\theta$ and $\hat\phi$ of the two vector boson decay products are measured in the $\ell\bar\ell$ c.o.m. system~\cite{Banerjee:2019twi}.}\label{fig:sketch}
\end{figure}

For testing the algorithm, we use different models of a Higgs boson~(\Ph) decaying as $\Ph\rightarrow\cPqb\bar\cPqb$ in association with a leptonically decaying \PZ boson.
Figure~\ref{fig:sketch} shows the decay planes of the $\textrm{p}\textrm{p}\rightarrow\PZ\Ph\rightarrow\cPqb\bar\cPqb\ell\bar\ell$ process. 
The angle $\Theta$ between the pp beam axis and the $\PZ\Ph$ system and the angle $\hat\phi$ of the \PZ decay plane are measured in the $\PZ\Ph$ center-of-mass (c.o.m.) system. 
The angle $\hat\theta$ of the two leptons~($\ell=e,\mu$) from the \PZ decay is measured in the $\ell\bar\ell$ c.o.m. system~\cite{Banerjee:2019twi}.

Associate $\PZ\Ph$ or $\PW\Ph$ production has been studied in the context of anomalous coupling approaches~\cite{Nakamura:2017ihk}, within the SM-EFT context~\cite{Banerjee:2019pks,Banerjee:2019twi}, as benchmark for neural-network based algorithms~\cite{Brehmer:2019gmn}, and the ATLAS and CMS Collaborations have measured kinematic properties in the different final states~\cite{ATLAS:2020fcp,ATLAS:2020jwz,CMS:2016tad}.
We start with a completely tractable toy model at the ME level, providing a background-free testbed where we can compare the performance of the BIT test statistic to the analytically known optimum.

\subsection{An analytic toy model of the \texorpdfstring{$\PZ\Ph$}{Zh} process}\label{sec:analytic-toy}

The LO SM-EFT modifications of the $\PZ\Ph$ process at the amplitude level have been computed in  Ref.~\cite{Banerjee:2019twi} and the formulae for differential cross section for $\textrm{pp}\rightarrow\PZ\Ph$ are found in Ref.~\cite{Nakamura:2017ihk}. 
We restrict to the $\PZ\rightarrow \ell \bar\ell$ decay channels and consider the Lagrangian in Eq.~\ref{eq:SMEFT-Lagrangian} with the dimension-6 SM-EFT operators
\begin{align}
\mathcal{O}_{\PH\PW}&=H^\dagger H\,W^i_{\mu\nu}W^{i\mu\nu},\label{eq:OHW}\\
\mathcal{O}_{\PH\widetilde{\PW}}&=H^\dagger H\,\widetilde{W}^i_{\mu\nu}W^{i\mu\nu},\label{eq:OHWtil}\\
\mathcal{O}_{\PH\textrm{Q}^{(3)}}&=iH^\dagger\te{D}^i_\mu H\,\bar Q\sigma^i\gamma^\mu Q.\label{eq:OHQ3}
\end{align}
The Wilson coefficients $\boldsymbol{\theta}=(\CHW,\CHWtilde,\CHQ)$ span the three-dimensional parameter space. 
A nonzero \CHW modifies the CP conserving $\Ph$--$\PZ$ coupling, while \CHWtilde parameterizes CP-violating effects. 
These modifications introduce energy growth via the derivative couplings in the $\textrm{SU}(2)$ field strength $W^i_{\mu\nu}$, and it's dual $\widetilde W^i_{\mu\nu}$, while the effects of the current-current interaction controlled by \CHQ grow with energy because of the $\PZ\Ph\bar q q$ four-point contact interaction. 

The analytic prediction for the MEs for $\PZ\Ph$ production, including the quadratic SM-EFT terms deliberately omitted in Ref.~\cite{Banerjee:2019twi}, is  
\begin{align}
\mathcal{M}^{\lambda=\pm}_\sigma(q \bar q)&=\sigma\frac{1+\sigma\lambda \cos\Theta}{\sqrt{2}}\hat{M}^{\lambda=\pm}_\sigma,\nonumber\\
\mathcal{M}^{\lambda=0}_\sigma(q \bar q)&=\sin\Theta\hat{M}^{\lambda=0}_\sigma,\nonumber\\
\mathcal{M}^{\lambda=\pm}_\sigma(\bar q q)&=-\sigma\frac{1-\sigma\lambda \cos\Theta}{\sqrt{2}}\hat{M}^{\lambda=\pm}_\sigma,\nonumber\\
\mathcal{M}^{\lambda=0}_\sigma(\bar q q)&=\sin\Theta\hat{M}^{\lambda=0}_\sigma,
\end{align}
with the functions
\begin{align}
\hat{\mathcal{M}}^{\lambda=\pm}_\sigma &=g_\PZ m_\PZ\sqrt{\hat s}\Bigg[\frac{g_{\PZ\sigma}}{\hat s-m_\PZ^2}+c_{\theta_W}\left(1+\frac{\hat s-m^2_\Ph}{m^2_\PZ}\right)\left(\frac{g_{\PZ\sigma}c_{\theta_W}}{\hat s-m_\PZ^2}+\frac{Q_q es_{\theta_W}}{\hat s}\right)\frac{v^2}{\Lambda^2}C_{\PH\PW}\nonumber\\\label{eq:Mpm}
&-\frac{2i\lambda k\sqrt{\hat s}}{m^2_\PZ}c_{\theta_W}\left(\frac{g_{\PZ\sigma}c_{\theta_W}}{\hat s-m_\PZ^2}+\frac{Q_q e s_{\theta_W}}{\hat s}\right)\frac{v^2}{\Lambda^2}C_{\PH\widetilde\PW}\Bigg]+g_\PZ^2\frac{\sqrt{\hat s}}{m_\PZ}T_q^{(3)}\frac{v^2}{\Lambda^2}\CHQ,
\end{align}
and
\begin{align}
\hat{\mathcal{M}}^{\lambda=0}_\sigma &=-g_\PZ w\sqrt{\hat s}\Bigg[\frac{g_{\PZ\sigma}}{\hat s-m^2_\PZ}\nonumber\\
&+c_{\theta_W}\left(1+\frac{\hat s-m^2_\Ph}{m^2_\PZ}-\frac{2k^2\sqrt{\hat s}}{m^2_\PZ w}\right)\left(\frac{g_{\PZ\sigma}c_{\theta_W}}{\hat s-m_\PZ^2}+\frac{Q_q e s_{\theta_W}}{\hat s}\right)\frac{v^2}{\Lambda^2}C_{\PH\PW}\Bigg]\nonumber\\
&-g^2_\PZ T^{(3)}_q\frac{w\sqrt{\hat s}}{m^2_\PZ}\frac{v^2}{\Lambda^2}\CHQ.\label{eq:M0}
\end{align}
The \PZ-boson polarisation $\lambda$ and the quark helicity $\sigma$ can take the values $\lambda=0,\pm1$ and $\sigma=\pm1$, respectively. 
The last terms in Eqs.~\ref{eq:Mpm}--\ref{eq:M0} are the contributions from the $\PZ\Ph\bar q q$ contact interaction which are not included in Ref.~\cite{Nakamura:2017ihk}, but are obtained from the replacement\footnote{The translation of the SM-EFT contributions considered in this work and described in Ref.~\cite{Banerjee:2019twi} into the conventions of Ref.~\cite{Nakamura:2017ihk} is obtained by $b_\PW=\kappa_{\PW\PW}/2$, $b_\PZ=\kappa_{\PZ\PZ}/2$,  $\tilde b_\PW=\tilde\kappa_{\PW\PW}/2$,  $\tilde b_\PZ=\tilde\kappa_{\PZ\PZ}/2$, $b_\gamma=\kappa_{\PZ\gamma}$, $\tilde b_{\gamma}=\tilde\kappa_{\PZ\gamma}$ and the contact interactions are implemented by Eq.~\ref{eq:contactZh}.} 
\begin{equation}
g_{\PZ\sigma}\rightarrow g_{\PZ\sigma} +g_{\PZ\sigma} g_\PZ  \frac{v^2}{\Lambda^2}\frac{\hat s-m_\PZ^2}{m^2_{\PZ}} T^{(3)}_q \CHQ\label{eq:contactZh}
\end{equation}
in the first lines in Eqs.~\ref{eq:Mpm}--\ref{eq:M0}.
The coupling constants are given by $g_\PZ=g/c_{\theta_W}$, $g_{\PZ+}=-g_\PZ Q_q s^2_{\theta_W}$, and $g_{\PZ-}=g_\PZ (T^{(3)}_q-Q_q s^2_{\theta_W})$ where $T^{(3)}_u=1/2$, $T^{(3)}_d=-1/2$, and $Q_q$ is the electric charge of the initial quark $\textrm{q}$, that is any of $\textrm{u}$, $\textrm{d}$, $\textrm{c}$, $\textrm{s}$, or $\textrm{b}$. 
The sine (cosine) of the Weinberg angle is denoted by $s_{\theta_W}$~($c_{\theta_W}$), the mass of the \PZ boson by $m_\PZ$, $v=246$\GeV is the vacuum expectation value of the Higgs field, and $g$ is the SM weak isospin coupling. 
To simplify the expressions, we use the \PZ boson energy $w=(\hat s+m^2_\PZ-m^2_\PH)/(2\sqrt{\hat s})$ and momentum $k=\sqrt{w^2-m^2_\PZ}$, that are given in terms of the squared center-of-mass energy~($\hat s$) of the $\PZ\Ph$-system~\cite{Nakamura:2017ihk}. 
The Wigner functions for the lepton helicity $\tau=\pm1$ are
\begin{align}
d_{\lambda=\pm}^\tau&=\;\tau\frac{1+\lambda\tau\cos\hat\theta}{\sqrt{2}}e^{i\lambda\hat\phi}\;\;\textrm{and}\nonumber\\
d_{\lambda=0}^\tau&=\;\sin\hat\theta.
\end{align}
The density matrices are
\begin{align}
\sum_\sigma\rho^{\lambda'\lambda}_\sigma(q\bar q)&=\{\mathcal{M}_\sigma^{\lambda'}(q\bar q)\}^\ast\mathcal{M}_\sigma^{\lambda}(q\bar q)\;\;\textrm{and}\nonumber\\
\sum_\sigma\rho^{\lambda'\lambda}_\sigma(\bar q q)&=\{\mathcal{M}_\sigma^{\lambda'}(\bar q q)\}^\ast\mathcal{M}_\sigma^{\lambda}(\bar q q),
\end{align}
and the differential cross section for $\textrm{pp}\rightarrow\PZ\Ph\rightarrow\ell\bar\ell\,\Ph$~\cite{Nakamura:2017ihk} in terms of the density matrices is
\begin{align}
&\frac{\textrm{d}\sigma^{\PZ\Ph}}{\textrm{d}\hat s\,\textrm{d}y\,\textrm{d}\cos\theta\,\textrm{d}\cos\hat\theta\,\textrm{d}\hat\phi}=\frac{m_\PZ k}{12288\pi^3\Gamma_\PZ s \hat s^{3/2}}\times\nonumber\\&\sum_f\sum_\tau|g_{\PZ\ell\bar \ell}^\tau|^2\sum_{q=u,d,c,s,b}\Bigg(q(x_1)\bar q(x_2)d^{\tau\dagger}\sum_\sigma\rho_\sigma(q\bar q)d^\tau+\bar q(x_1) q(x_2)d^{\tau\dagger}\sum_\sigma\rho_\sigma(\bar q q)d^\tau\Bigg),\nonumber\\\label{eq:diff-xsec-Zh}
\end{align}
where the $\PZ$--$\ell$ couplings are given by $g^+_{\PZ\ell\bar\ell}=g_Z s_{\theta_W}^2\phantom{\Big|\;\;}$ and $g^-_{\PZ\ell\bar\ell}=g_Z (-1/2+s_{\theta_W}^2)$ and $\tau=\pm$ denotes the lepton helicity. 
For calculating the parton luminosities, we extract the nCTEQ15~\cite{Kusina:2015vfa} parton distribution functions $q(x)$ and $\bar q(x)$ from the \textsc{ManeParse} package~\cite{Clark:2016jgm}.

We implement Eq.~\ref{eq:diff-xsec-Zh} in a toy simulation, including the second order in the Wilson coefficients. The variable $\hat s$ is sampled according to the parton luminosities, and all other variables are sampled uniformly. 
We focus the study on the highly energetic tails of the kinematic distributions and implement a requirement of $p_{\textrm{T},\textrm{Z}}>200$~\GeV on the transverse momentum of the \PZ boson at the generation level. 
For each event, the weight functions from Eq.~\ref{eq:wi-poly-expansion} are computed for the choice $\boldsymbol{\theta}_0=0$, i.e., we simulate at the SM parameter point. 
An important simplifcation of the toy study is the absence of detector simulation and the latent variables, such that formally $p(\boldsymbol{x}|\boldsymbol{z})=\delta(\boldsymbol{x}-\boldsymbol{z})$. All toy likelihood functions are, thus, fully tractable.

\begin{figure}\centering
        \includegraphics[width=0.49\textwidth]{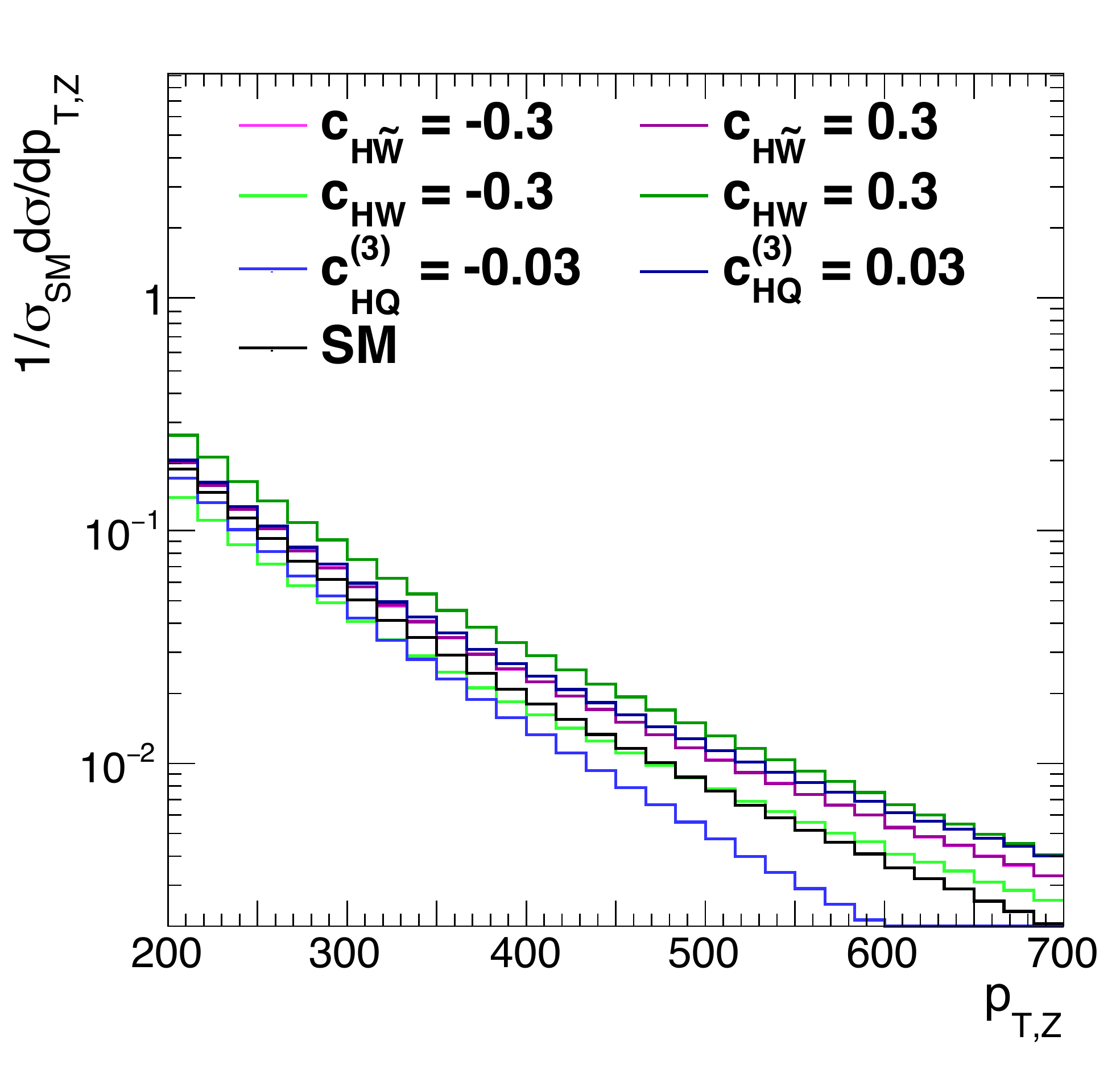}
        \hfill
        \includegraphics[width=0.49\textwidth]{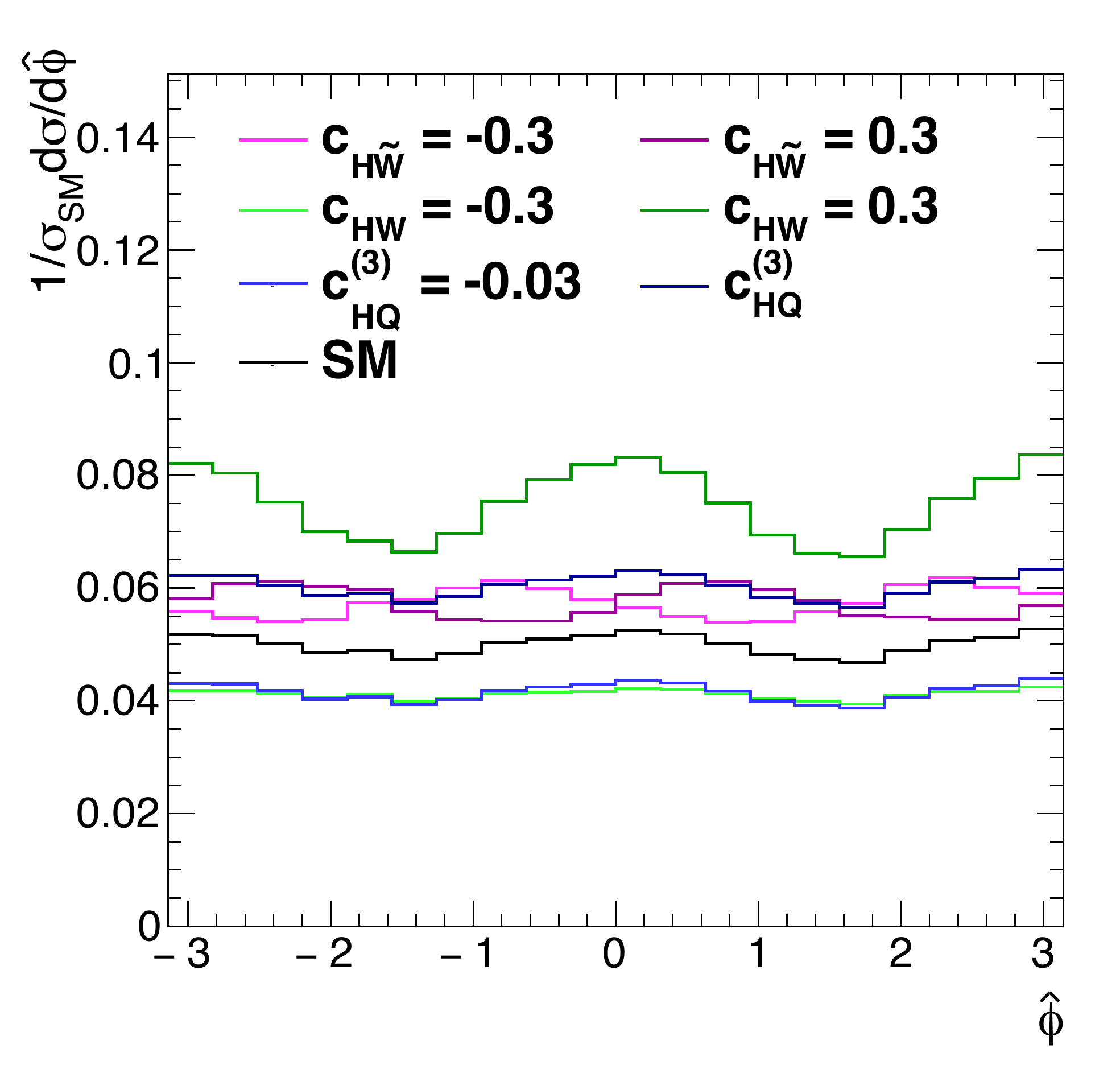}
    \caption{
    The normalized differential distributions $1/\sigma_{\boldsymbol{\theta}_\textrm{SM}}\textrm{d}\sigma/\textrm{d}p_{\textrm{T},\textrm{Z}}$ and $1/\sigma_{\boldsymbol{\theta}_\textrm{SM}}\textrm{d}\sigma/\textrm{d}\hat\phi$ for various values of the Wilson coefficients in the analytic model in Eq.~\ref{eq:diff-xsec-Zh}. The distributions are normalized at $\boldsymbol{\theta}_\textrm{SM}=0$ such that the dependence of the normalization on $\boldsymbol{\theta}$ is visible.} \label{fig:features}
\end{figure}
We check the implementation by comparing the predicted differential cross section from these weight functions using Eq.~\ref{eq:diff-xsec-approx} with independently obtained simulated samples at various values of $\boldsymbol{\theta}\neq\boldsymbol{\theta}_\textrm{SM}$. 
We find agreement within the statistical precision of the simulation. 
As examples, we show the distribution of $p_{\textrm{T},\textrm{Z}}$ and $\hat\phi$ for various values of the Wilson coefficients in Fig.~\ref{fig:features}. 
The normalized differential cross section in $p_{\textrm{T},\textrm{Z}}$ in Fig.~\ref{fig:features}~(left) indicates that SM-EFT induced shape effects appear in the highly energetic tails, in particular for \CHQ. Figure~\ref{fig:features}~(right) shows the sensitivity of the $\hat\phi$ distribution to non-zero values of $C_{\PH\PW}$ and $C_{\PH\widetilde{\PW}}$.

\subsection{Learning from the toy model}

\begin{figure}\centering
        \includegraphics[width=0.48\textwidth]{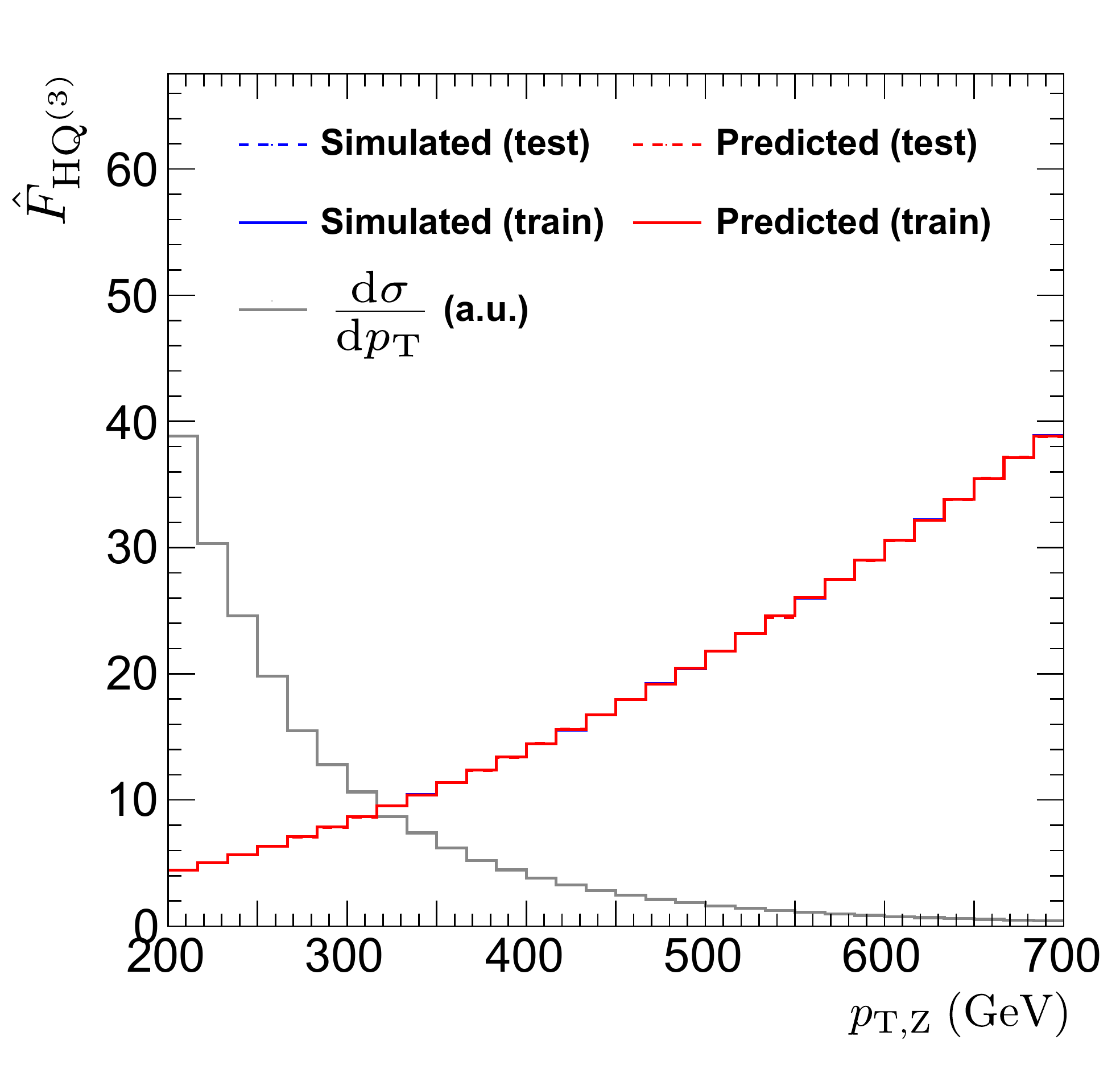}
        \hspace{-.2cm}\includegraphics[width=0.495\textwidth]{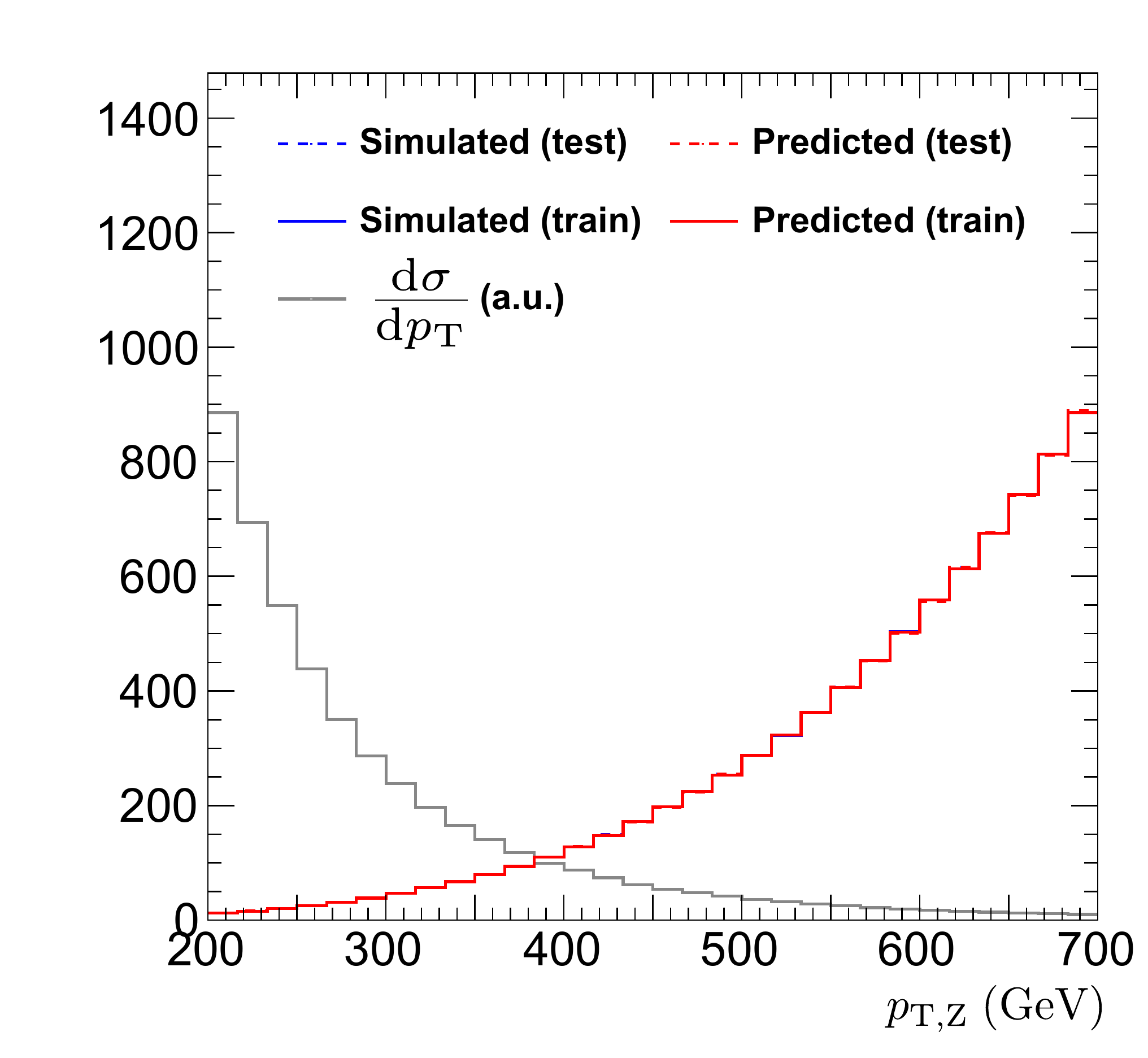}\\
        \includegraphics[width=0.48\textwidth]{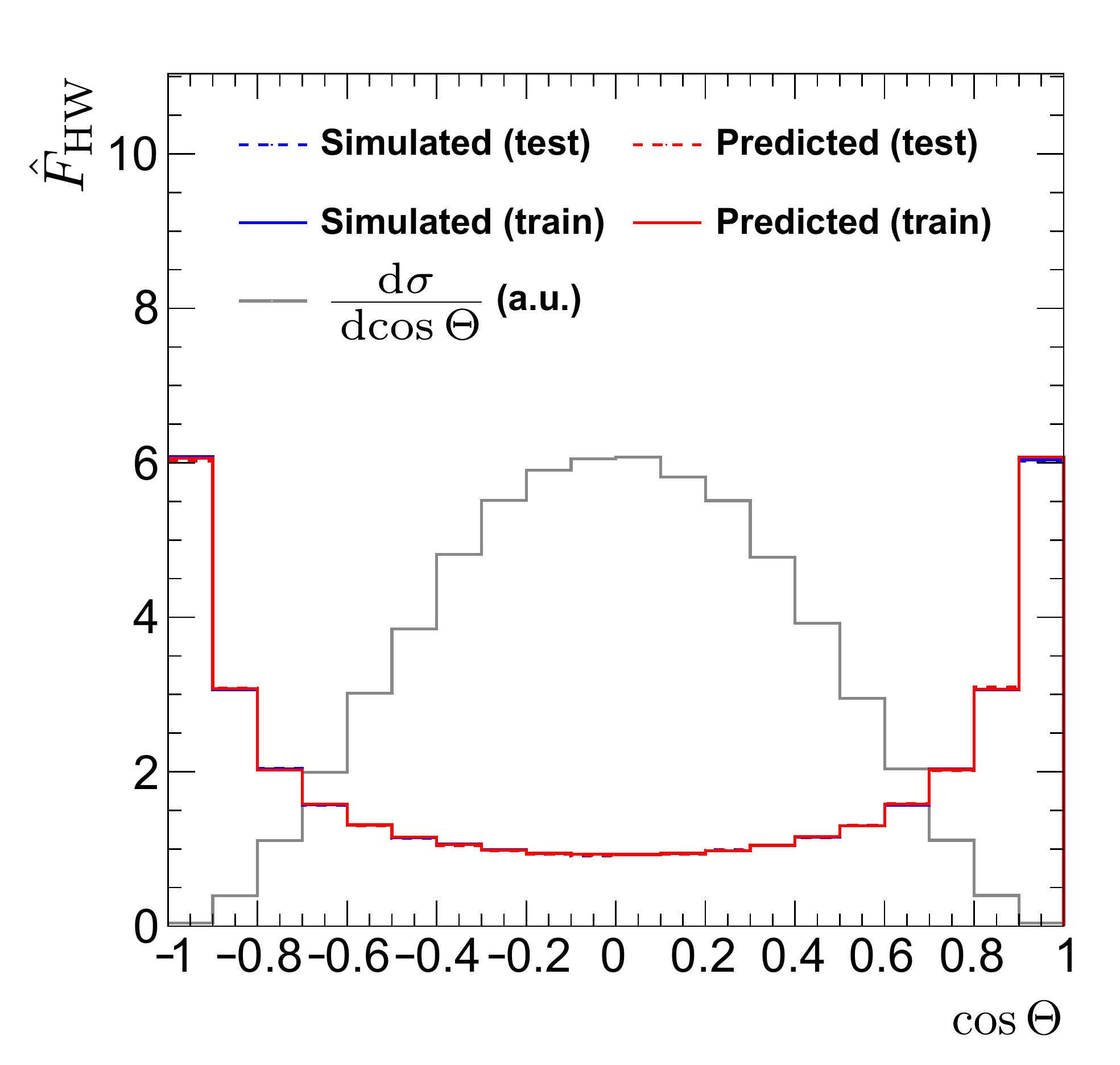}
        \includegraphics[width=0.48\textwidth]{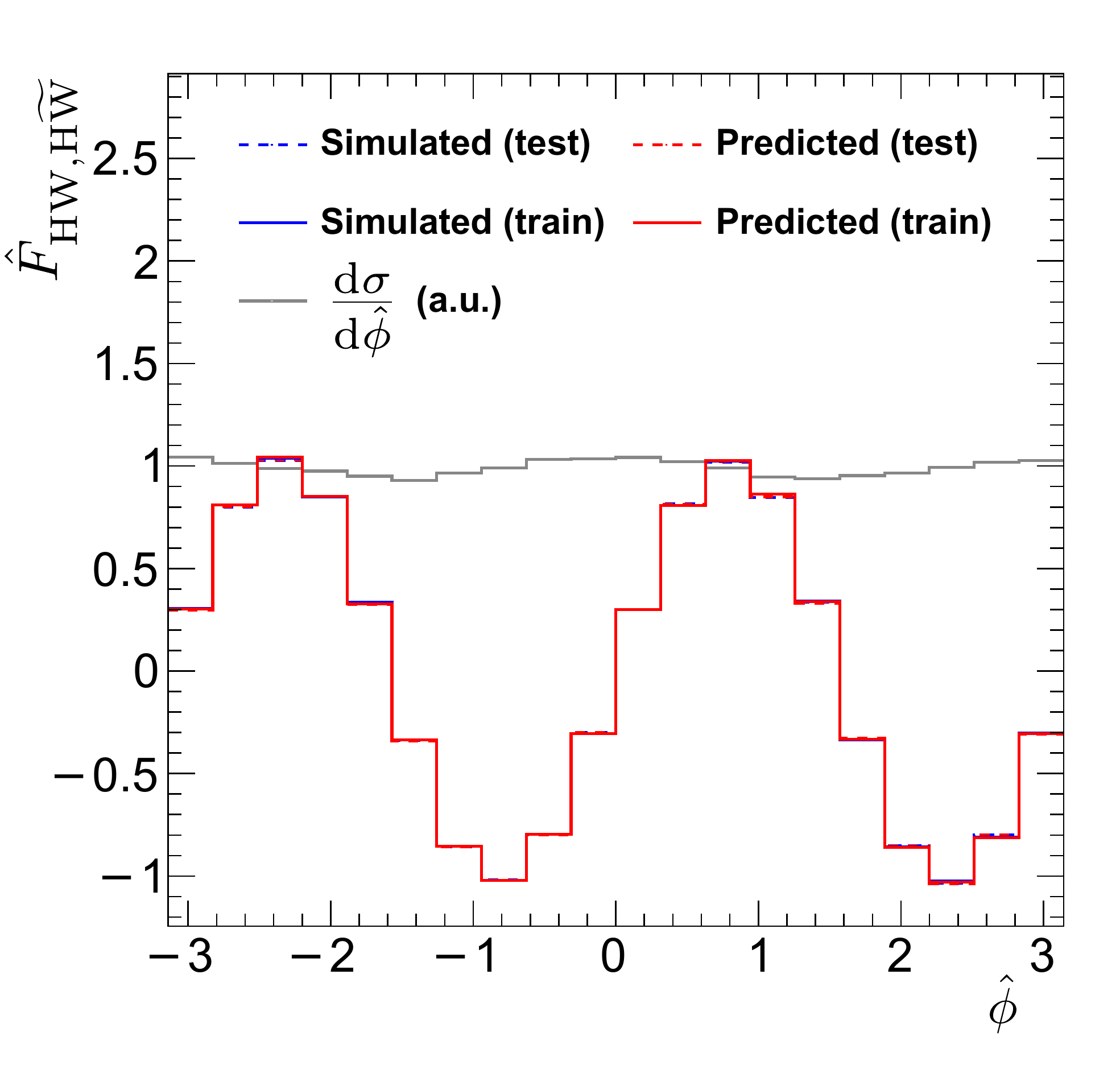}
    \caption{Predicted and true mean values of the linear coefficient function $\hat F_{\PH\PQ^{(3)}}$~(top left) and the second-order coefficient function $\hat F_{\PH\PQ^{(3)},\PH\PQ^{(3)}}$~(top right) as a function of $\textrm{p}_{T,\PZ}$.  The linear coefficient function $\hat F_{\PH\PW}$~(bottom left) is provided as a function of $\cos\Theta$ and the mixed second-order coefficient function $\hat F_{\PH\PW,\PH\widetilde{\PW}}$~(bottom right) is shown as a function of $\hat\phi$. The simulated true values are shown in blue color, while the predictions from the estimators are shown in red color. The training data set is used to obtain the solid lines, while the statistically independent test data set is shown with dashed lines. For illustration, the mean of the differential cross section of the observable on the $x$-axis is also shown in gray color with arbitrary normalization.}\label{fig:scores}
\end{figure}

Next, we train BITs using our \textsc{Python} implementation of Algorithm~\ref{algo:bit} from Ref.~\cite{CHATTERJEE2022108385} with $N_\textrm{sim}=2\cdot10^6$. We choose $B=250$, $N_{\textrm{min}}=50$, and a maximum tree depth of $D=5$ for training estimators $\hat F_{a}$ and $\hat F_{ab}$ for the~9~coefficient functions $R_a$ and $R_{ab}$.
The three coefficient functions $R_{a}$ correspond to the linear dependence on $C_{\PH\PQ^{(3)}}$, $C_{\PH\PW}$, and $C_{\PH\widetilde{\PW}}$, and the six second-order coefficient functions $R_{ab}$ correspond to the quadratic terms in these Wilson coefficients, including the mixed quadratic terms.
The training variables are 
\begin{equation}
    p_{\textrm{T,Z}}, \;y, \; \Theta, \;\hat\theta, \;\hat\phi 
\end{equation}
and, to slightly improve the speed of convergence, the redundant angular observables $f_{LL}$, \ldots, $\tilde f_{TT'}$ as defined in Ref.~\cite{Banerjee:2019twi}. 
For the sake of brevity, the resulting estimators of the linear and quadratic dependence are labeled as $\hat F_{\PH\PW}(\boldsymbol{x})$, $\hat F_{\PH\PW,\PH\PW}(\boldsymbol{x})$, $\hat F_{\PH\PW,\PH\widetilde\PW}(\boldsymbol{x})$, \etc
As examples, the predicted linear coefficient function $\hat F_{\PH\PQ^{(3)}}$ and the second-order coefficient function $\hat F_{\PH\PQ^{(3)},\PH\PQ^{(3)}}$ are shown as a function of $\textrm{p}_{T,\PZ}$ in Fig.~\ref{fig:scores}.
Furthermore, the linear coefficient function $\hat F_{\PH\PW}$ is provided as a function of $\cos\Theta$ and the mixed second-order coefficient function $\hat F_{\PH\PW,\PH\widetilde{\PW}}$ is shown as a function of $\hat\phi$. 
The value on the $y$-axis corresponds to the mean as obtained from the toy simulation and integrated over all other observables. 
The large values for the linear and quadratic dependence on \CHQ reflect the sensitive dependence on this Wilson coefficient, in particular in the tails of $p_{\textrm{T,Z}}$. All distributions are obtained for the training sample and a statistically independent test sample. Furthermore, the predicted cross-section coefficients are compared to the corresponding true values, which can easily be obtained from Eq.~\ref{eq:R-poly-expansion} in this simple toy study. 
A near-perfect agreement is found in all cases. For illustration, the distribution of the observable on the $x$-axis is overlayed in gray color with arbitrary normalization. 
It shows that phase space tails with a large dependence on a Wilson coefficient can be sparsely populated.

\subsection{Hypothesis tests}
We use the BIT estimator in non-compound~(simple) hypothesis tests~\cite{Cranmer:2014lly} and compare its performance to the theoretically achievable optimum.
Given a generic statistic $t_{\boldsymbol{\theta}}$ that may depend on $\boldsymbol{\theta}$ parametrically, a value for the integrated luminosity $\mathcal{L}$, and the analytic model for $\PZ\Ph$ production in Eq.~\ref{eq:diff-xsec-Zh}, we generate a large number of random toy data sets $\{\mathcal{D}\}$, each containing a number of events that is itself Poisson distributed with mean $\mathcal{L}\sigma(\boldsymbol{\theta})$. 
In this way, the toys sample both the observed number of events and the kinematic distributions for a given set of Wilson coefficients. 
For the exclusion of SM-EFT parameter space, the null hypothesis is a parameter point $\boldsymbol{\theta}\neq\boldsymbol{\theta}_\textrm{SM}$. 
The alternate hypothesis is $\boldsymbol{\theta}_0=\boldsymbol{\theta}_\textrm{SM}=0$. 
The probability of rejecting the null hypothesis when it is true~(type-1 error) is called the test's size, and we adopt the conventional choice of $5\%$.
The distribution of the test statistic under the null hypothesis $p(t_{\boldsymbol{\theta}}|\boldsymbol{\theta})$ is evaluated with toys and used to define the p-value of an observation as
\begin{equation}
\bar p(t_{\boldsymbol{\theta}},\boldsymbol{\theta})=\int_{t_{\boldsymbol{\theta}}}^\infty \textrm{d}t_{\boldsymbol{\theta}}'\, p(t_{\boldsymbol{\theta}}'|\boldsymbol{\theta}).
\end{equation}
It quantifies the consistency of the observation with the null hypothesis and is distributed uniformly between 0 and 1 under $\boldsymbol{\theta}$.
The median expected exclusion reach is defined by the implicit equation
\begin{equation}
  \bar p(t_{\textrm{med}}(\boldsymbol{\theta}_{95\%}),\boldsymbol{\theta}_{95\%})=0.05,\;\;\textrm{where}\;\;t_{\textrm{med}}(\boldsymbol{\theta}) = \textrm{Median}(t_{\boldsymbol{\theta}}|\boldsymbol{\theta}_0).
\end{equation}
The probability to accept the null hypothesis under the alternate  
\begin{equation}
    \beta=p(\bar p>0.05|\boldsymbol{\theta}_0)
\end{equation}
is the \mbox{type-2~error} probability, and $1-\beta$ is called the power of the test. 
In the absence of nuisance parameters, the optimal test statistic follows from the Neyman-Pearson lemma as the likelihood ratio $q_{\boldsymbol{\theta}}$, given in Eq.~\ref{eq:log-likelihood-general}. It has the highest power for a given test size.
In the toy simulation, we can use it as a reference for assessing the performance of the BIT test statistic $\hat q_{\boldsymbol{\theta}}$ in Eq.~\ref{eq:BIT-statistic}.

\begin{figure}\centering
        \raisebox{0mm}{\includegraphics[width=0.49\textwidth]{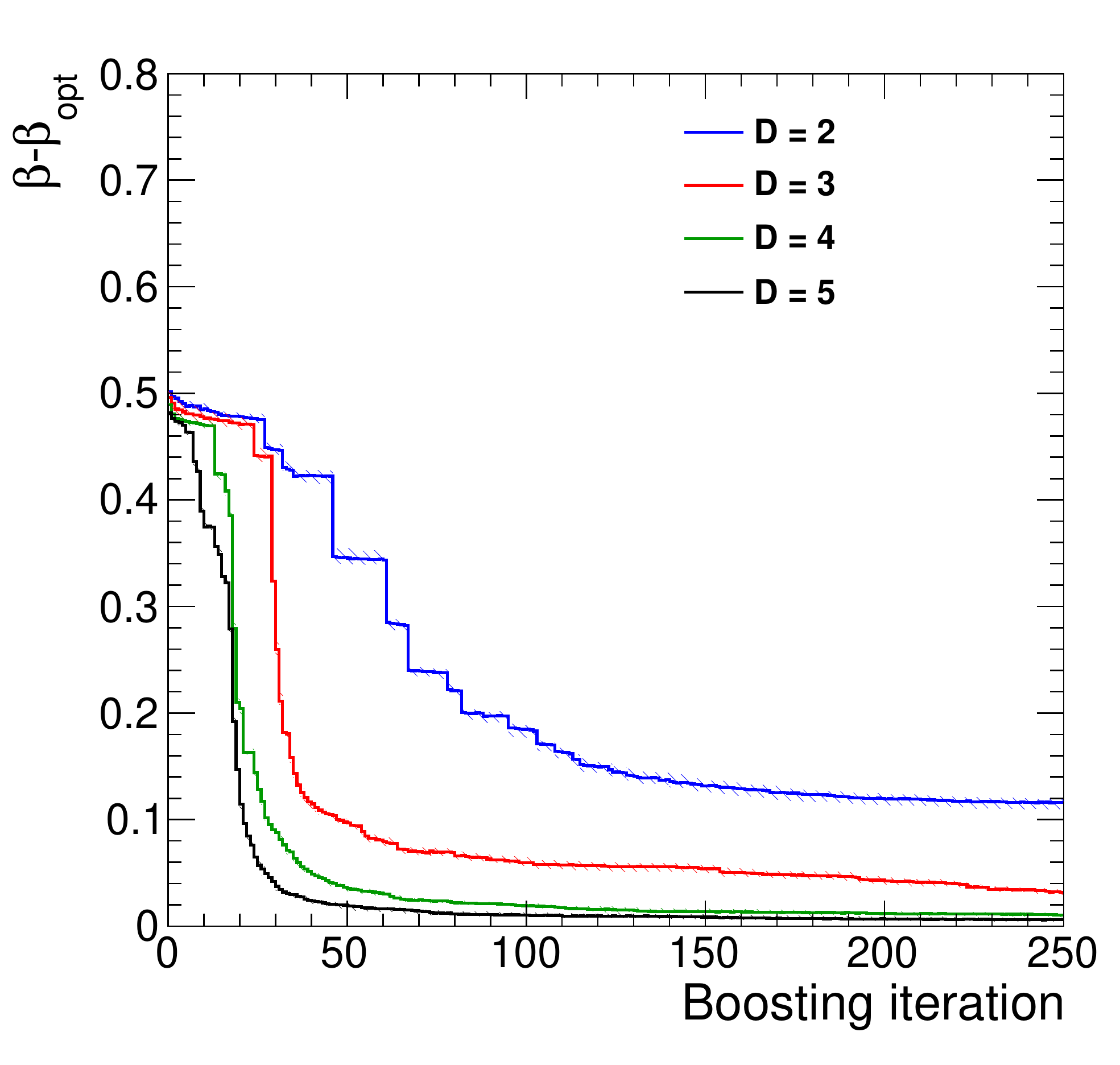}}
        \includegraphics[width=0.49\textwidth]{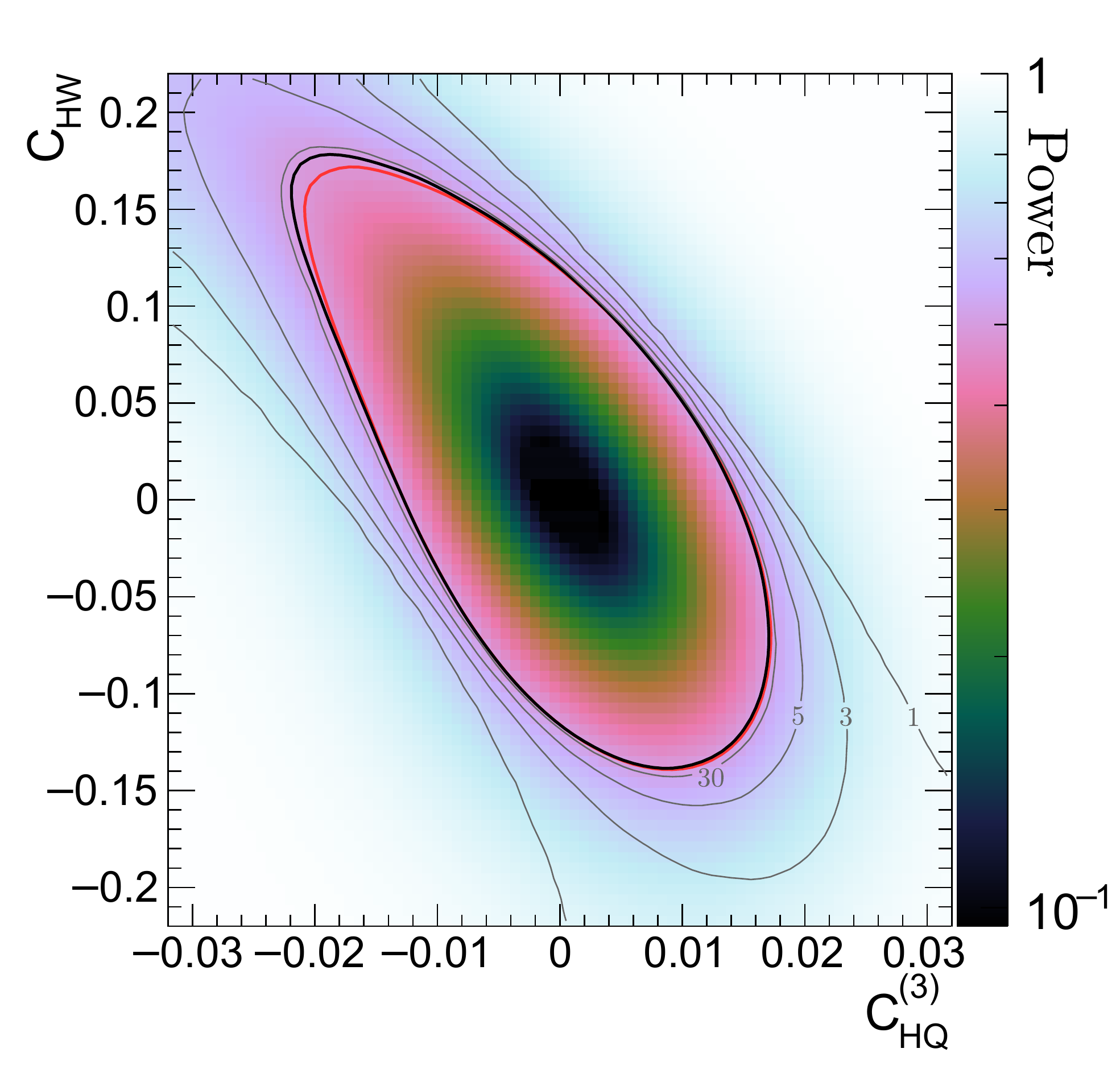}\\
        \includegraphics[width=0.49\textwidth]{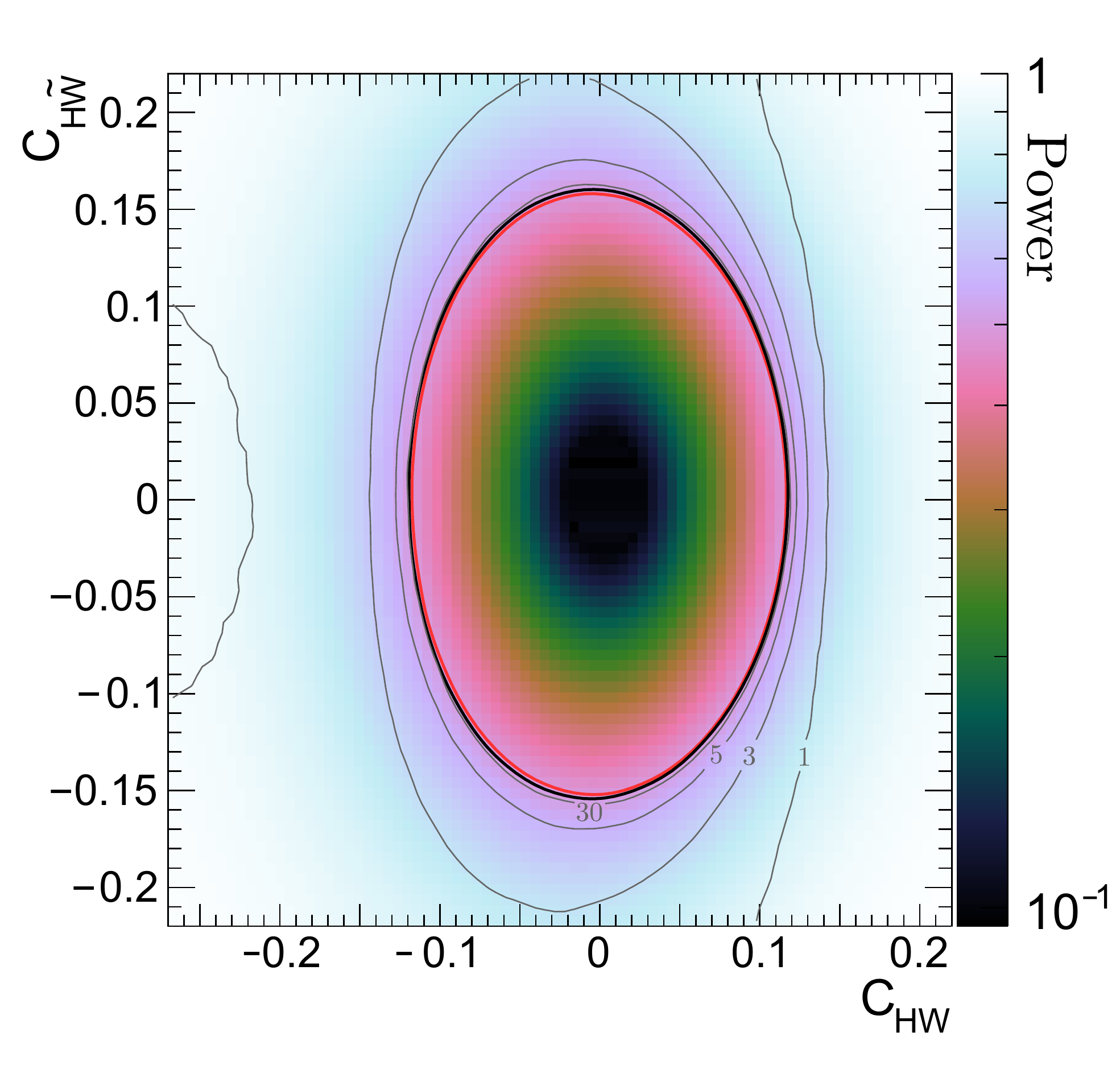}
        \includegraphics[width=0.49\textwidth]{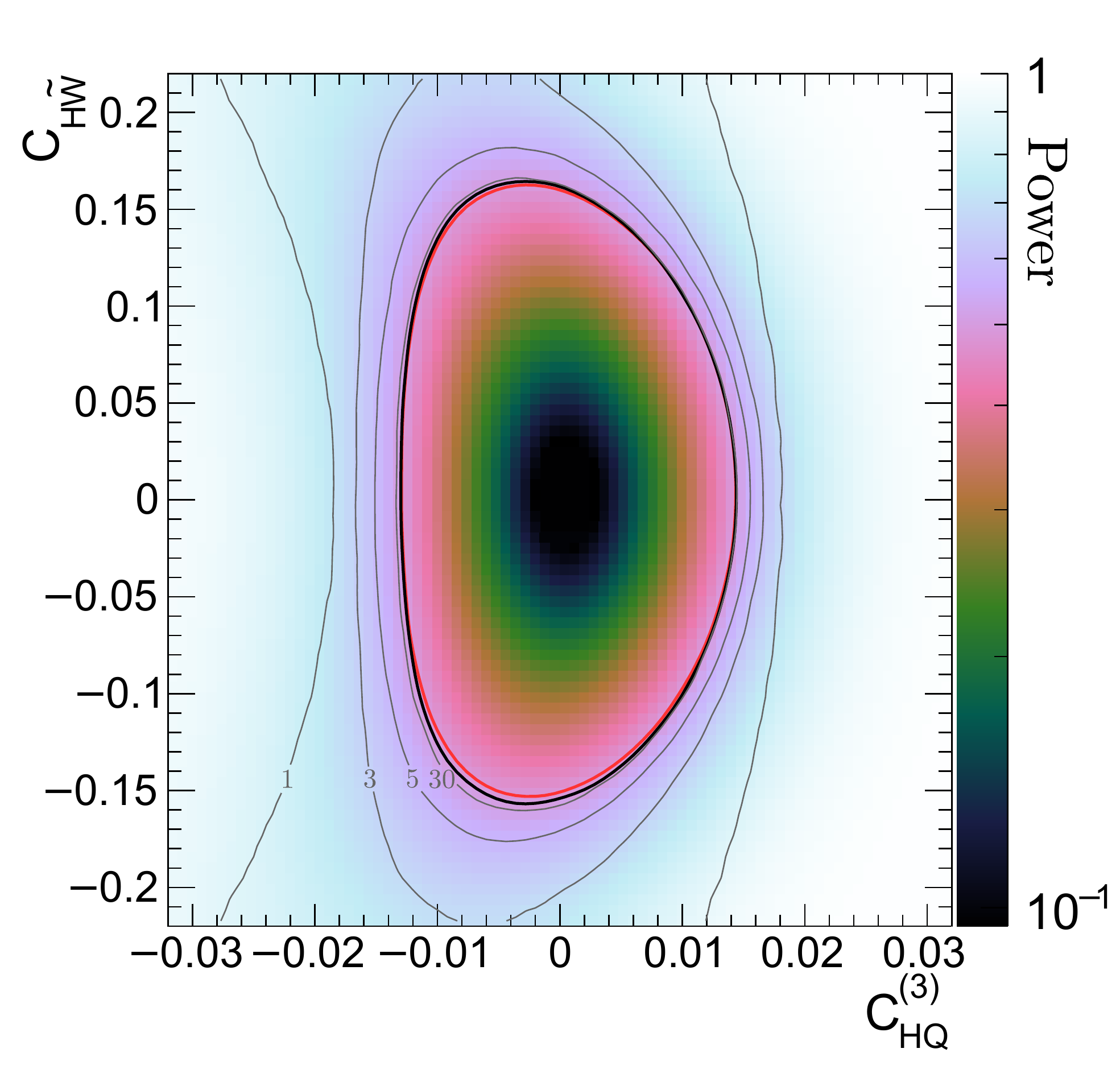}
    \caption{The evolution of the improvement in the power, $\beta-\beta_{\textrm{opt}}$, of the $\hat q_{\boldsymbol{\theta}}$ test statistic for $C_{\PH\widetilde{\PW}}=0.2$ as a function of the boosting iteration for different values of the maximum tree depths $D$~(top left). 
    Power of the $\hat q_{\boldsymbol{\theta}}$ test in the $C_{\PH\PQ^{(3)}}$--$C_{\PH\PW}$ plane~(top right), the $C_{\PH\PW}$--$C_{\PH\widetilde{\PW}}$ plane~(bottom left), and the $C_{\PH\PQ^{(3)}}$--$C_{\PH\widetilde{\PW}}$ plane~(bottom right). The red contours show the 95\%~CL median expected limit from the theoretically optimal test statistic $q_{\boldsymbol{\theta}}$, while the black contours correspond to the BIT estimate $\hat q_{\boldsymbol{\theta}}$. The contours shown in gray indicate the exclusion contours for different numbers of bins for the binned test statistic. }\label{fig:limits-toys-unbinned}
\end{figure}

To demonstrate the BIT's training convergence, we investigate the test's power as a function of the boosting iteration. 
We compute $\beta_\textrm{opt}$ for an arbitrary choice $C_{\PH\widetilde{\PW}}=0.2$ using the true likelihood, $q_{\boldsymbol{\theta}}$. 
Furthermore, we evaluate $\beta$ for the estimated unbinned BIT test statistic, $t_{\boldsymbol{\theta}}=\hat q_{\boldsymbol{\theta}}$, at each boosting iteration and for different maximum tree depths $D$.
Figure~\ref{fig:limits-toys-unbinned}~(top left) shows $\beta-\beta_{\textrm{opt}}$. 
Tree depths of 4 or 5 closely approach optimal performance, at least for this parameter point, justifying our earlier choice of $D=5$. 
Shallower trees lead to imperfect performance that are probably not improvable with a higher number of boosting iterations.
The shaded area shows the $1\sigma$ statistical uncertainty obtained from repeating this study 100 times. 
At the beginning of the training, when the BIT's performance is still poor, there is only the discrimination from the total yield that does not depend on the observed features. 
After about 200 iterations, the optimal performance is nearly reached, the full kinematic dependence is learned, and the power of the test improves by 46\% in this example.

\subsection{Exclusion contours and binned test statistic}\label{sec:toys-exclusion-binning}

Figure~\ref{fig:limits-toys-unbinned}~(top right and bottom row) also shows the power of the BIT estimate $\hat q_{\boldsymbol{\theta}}$ in the two-dimensional planes spanned by pairs of Wilson coefficients. The red contours show the median expected limit for the theoretical optimum $q_{\boldsymbol{\theta}}$, while the black lines correspond to the $\hat q_{\boldsymbol{\theta}}$. Near-perfect agreement is found for all exclusion contours, confirming near optimality of the BIT test statistic.

So far, we have dealt with unbinned test statistics. These are, however, the exception in LHC data analyses. 
Reasons include the large event counts, leading to prohibitive CPU consumption when, e.g., profiling nuisance parameters, and difficult experimental backgrounds estimated in discrete sidebands,  indirectly introducing binning effects. 
Therefore, we also study binned test statistics. 

For a chosen number of bins $M$, it is evident from, e.g., Fig.~\ref{fig:scores} that a fixed binning in  $R(\boldsymbol{x}|\boldsymbol{\theta},\boldsymbol{\theta}_0)$ is detrimental, because the relative importance of the linear and quadratic coefficients strongly depends on $\boldsymbol{\theta}$ and these functions can take vastly different values. 
A good binning choice should perform equally well in the whole parameter space and,  therefore, will in general depend on $\boldsymbol{\theta}$.  
The following simple algorithm solves this problem. 
For a given $\boldsymbol{\theta}$, we choose a binning such that the same fraction $1/M$ of signal events is expected in each bin of the distribution of $R$ under the null hypothesis~$\boldsymbol{\theta}$. This is achieved by transforming the toy distribution of $R$ to the unit interval and splitting it into $M$ equidistant bins. 
The resulting feature-space binning changes smoothly with $\boldsymbol{\theta}$.
In this way, the unbinned test statistic is approximated by the binned test statistic
\begin{equation}
    q_{\boldsymbol{\theta},\textrm{binned}}(\mathcal{D})=\sum_{j=1}^{M}\left(\lambda_j(\boldsymbol{\theta})-\lambda_j(\boldsymbol{\theta}_0)-n_j\log\frac{\lambda_j(\boldsymbol{\theta})}{\lambda_j(\boldsymbol{\theta}_0)}\right),\label{eq:binned-test-stat}
\end{equation}
where $\lambda_j$ is the Poisson mean of the yield in bin $j$ and $n_j$ is the number of observed events in this bin. 
It accumulates events with similar values of $R(\boldsymbol{x}|\boldsymbol{\theta},\boldsymbol{\theta}_0)$.
The quality of the approximation depends on the residual variation of $R(\boldsymbol{x}|\boldsymbol{\theta},\boldsymbol{\theta}_0)$ within the bin $j$.
The gray contours in Fig.~\ref{fig:limits-toys-unbinned}~(top right and bottom) correspond to different choices of $M$ for $\hat q_{\boldsymbol{\theta}}$. 
At $M=1$, all kinematic information is disregarded, and the test statistic corresponds to a Poisson counting experiment of the total yield. 
The median expected 95\% CL limit, therefore, is feeble.
The other contours are the 95\% CL limits for $M=2$,  $M=5$, and  $M=30$, and show a gradual improvement. 
For $M=30$, there is barely any difference between the binned and the unbinned test statistic, indicating that this value is sufficient.


\section{A \MGvATNLO model with backgrounds}\label{sec:binned}

Moving closer to realistic applications for LHC experiments,
we generate the $\PZ\Ph$ signal process and the most important background processes with \MGvATNLO~v2.6.0~\cite{Alwall:2014hca} and using the NNPDF parton distribution functions~v3.1~\cite{NNPDF:2017mvq}.
We simulate at a center-of-mass energy of $13\TeV$, followed by leptonic \PZ decays ($\ell=e,\mu,\tau$) and $\Ph\rightarrow\bar\cPqb\cPqb$. 
We use the \textsc{SMEFTsim} v3.0 model~\cite{Brivio:2020onw} which includes the operators in Eqs.~\ref{eq:OHW}--\ref{eq:OHQ3}. 
The ME simulation is interfaced to \PYTHIA v8.226 using the CP5 tune~\cite{Skands:2014pea,CMS:2015wcf,CMS:2019csb} for fragmentation, parton shower, and hadronization of partons in the initial and final states, along with the underlying event and the multiparton interactions.
We normalize the simulated sample to $\mathcal{L}=350\,$fb$^{-1}$. 
When correcting the differences in the parton luminosities, the model agrees within 10\%--20\% with the analytic toy model in Sec.~\ref{sec:analytic-toy} at the level of single-observable distributions. 

\begin{figure}\centering
        \raisebox{0mm}{\includegraphics[width=0.49\textwidth]{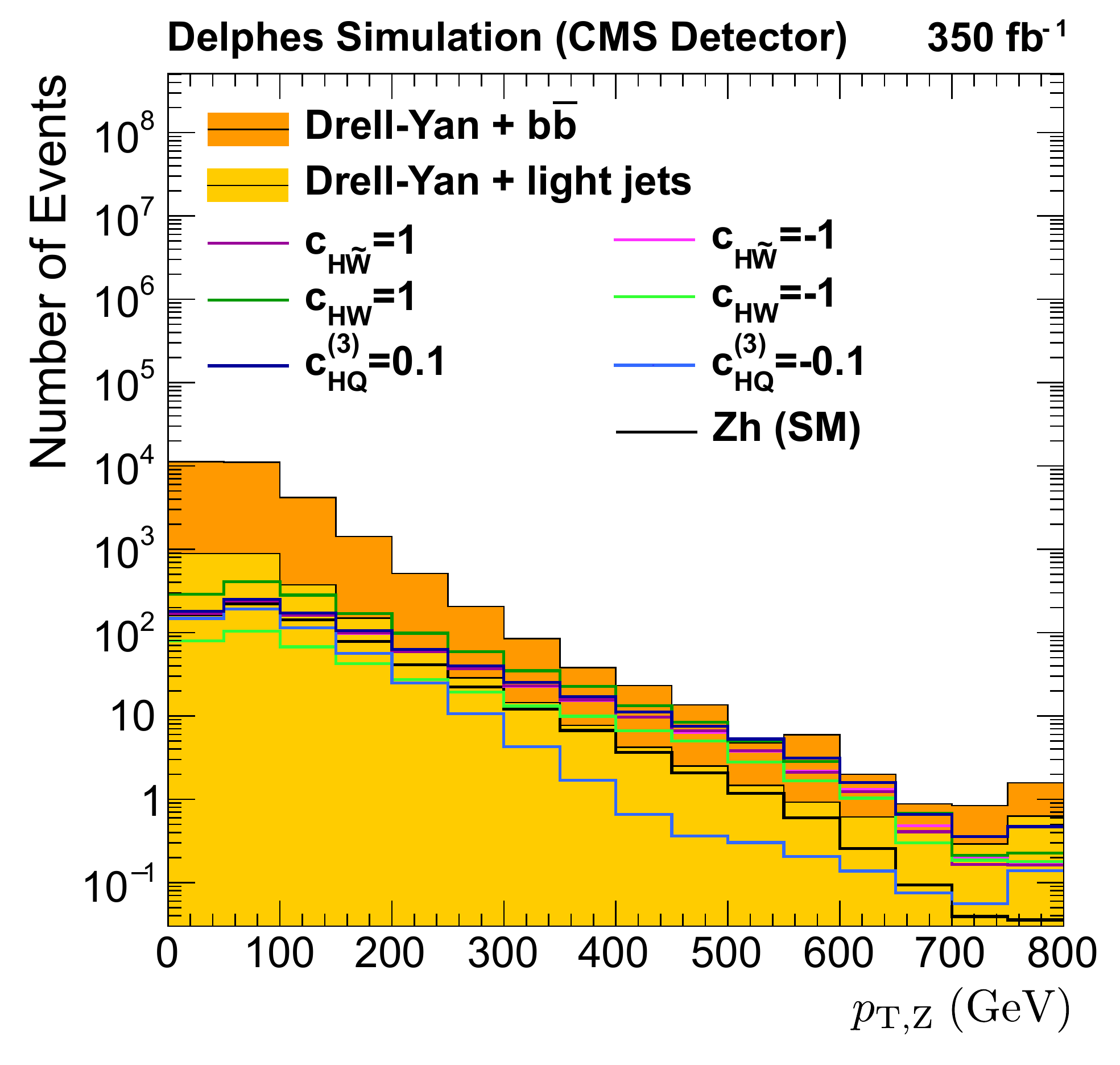}}
        \includegraphics[width=0.49\textwidth]{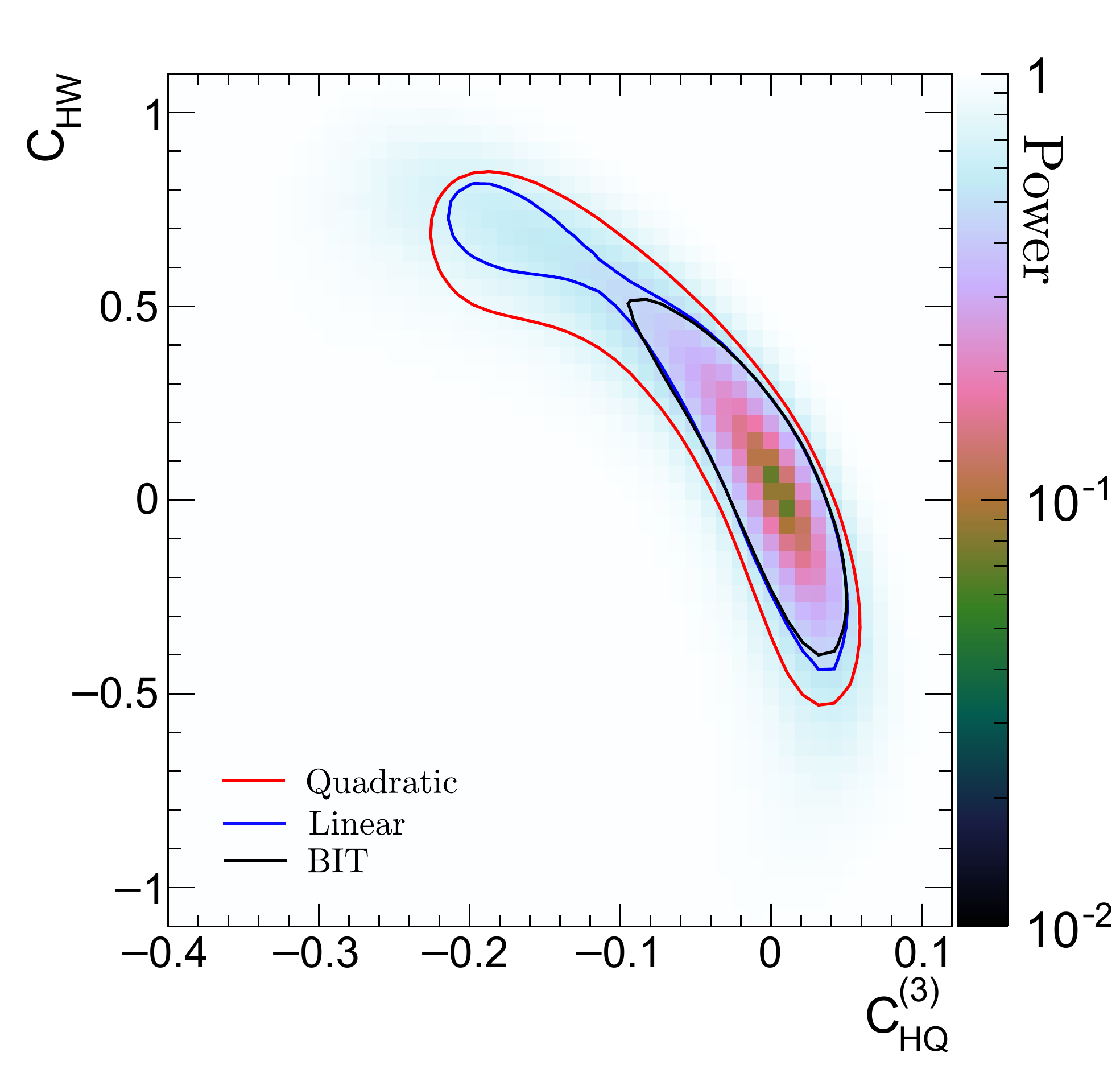}\\
        \includegraphics[width=0.49\textwidth]{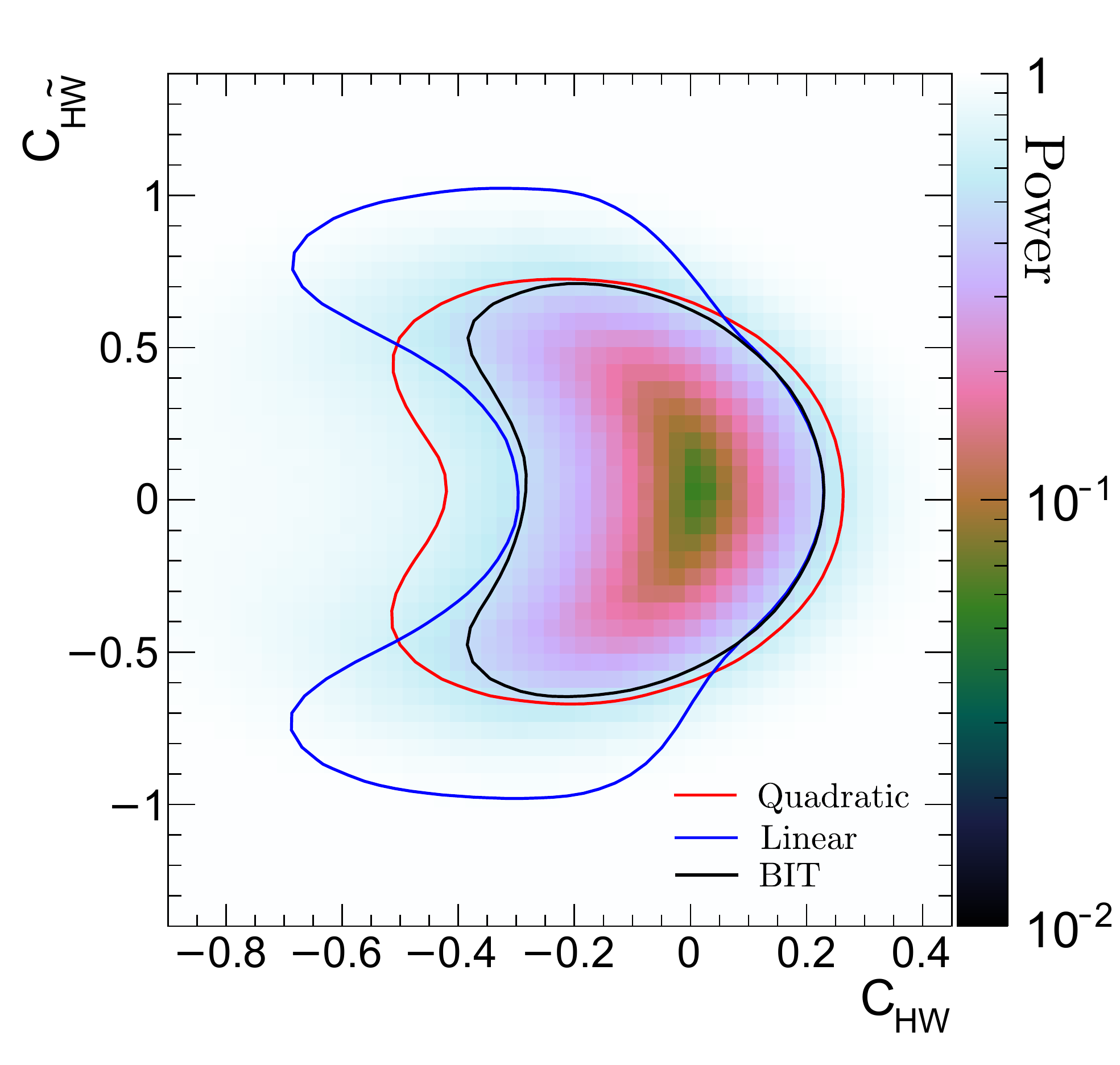}
        \includegraphics[width=0.49\textwidth]{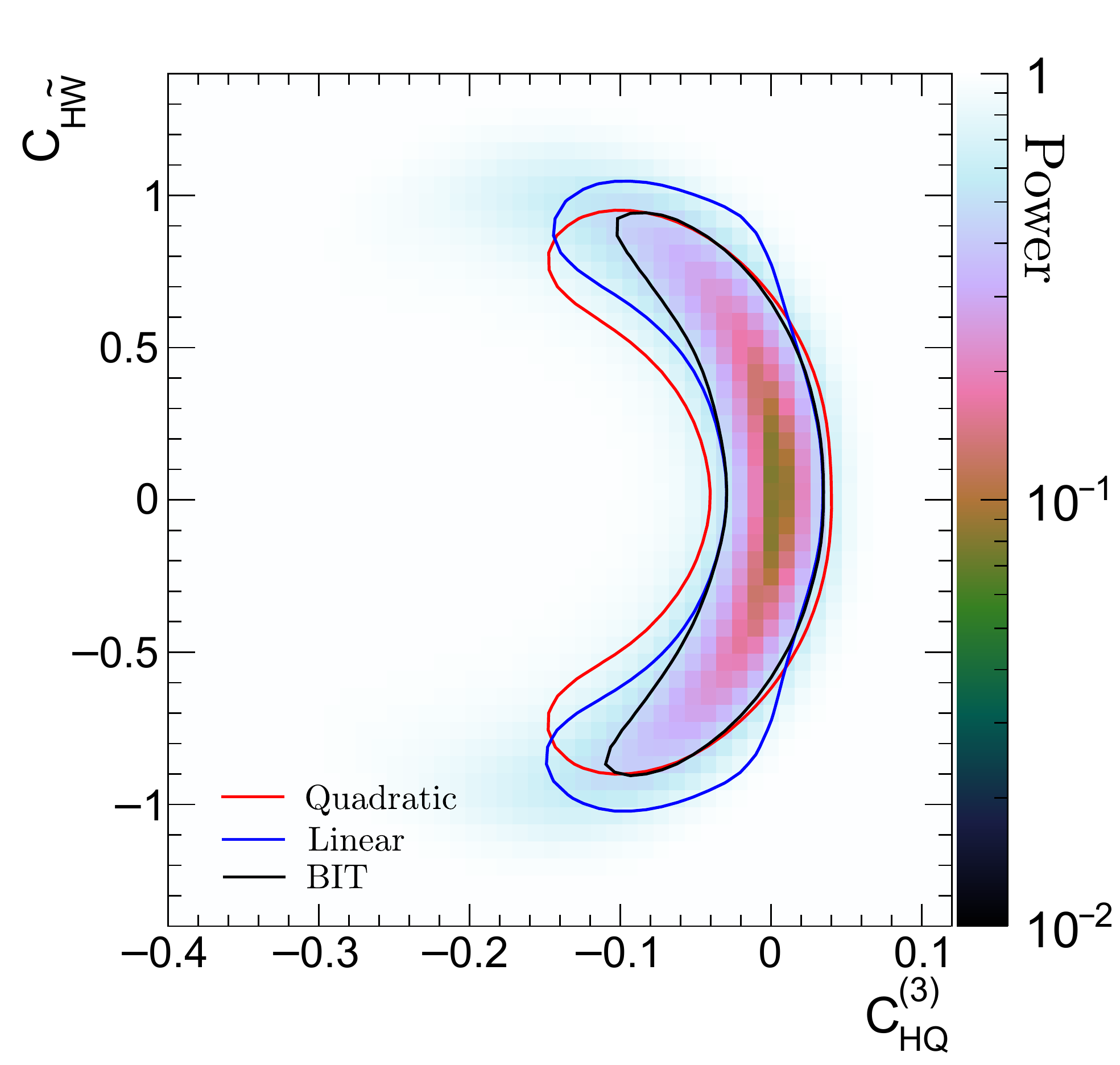}
    \caption{Spectrum of the transverse momentum of the \PZ~boson in  \Delphes simulation~(top left). The Drell--Yan background processes with (without) a pair of \cPqb~quarks at the generator level is shown in orange~(yellow).  The SM prediction for the $\PZ\Ph$ signal process is overlayed as black solid line, while other SM-EFT parameter points are shown in different colors. The last bin contains the overflow.
    Power of the $\hat q$ with $N_{\textrm{bins}}=30$ in the $C_{\PH\PQ^{(3)}}$--$C_{\PH\PW}$ plane~(top right), the $C_{\PH\PW}$--$C_{\PH\widetilde{\PW}}$ plane~(bottom left), and the  $C_{\PH\PQ^{(3)}}$--$C_{\PH\widetilde{\PW}}$ plane~(bottom right). The black contour shows the median expected limit from $\hat q$, while blue~(red) lines correspond to the exclusion contour obtained from the linear~(quadratic) terms as described in the text. }\label{fig:limits-delphes-binned}
\end{figure}

There are, nevertheless, the following important differences in the present study.  
The ME for the signal is simulated including up to one extra parton. 
Double counting of the partons generated with \MGvATNLO and \PYTHIA is removed using the MLM~\cite{Alwall:2007fs} scheme. 
The events are subsequently processed with a \Delphes~\cite{deFavereau:2013fsa}-based simulation model of the CMS detector, and kinematic cuts are placed on jets, electrons, and muons.
Jets are reconstructed with anti-kT algorithm\cite{Cacciari:2008gp} using a distance parameter of 0.4 in the \textsc{FastJet} software package\cite{Cacciari:2011ma}.
The \cPqb~tagging of jets is based on  parton-matching and a parametrization of the nominal CMS  \cPqb-tagging efficiency.
Electrons and muons must be isolated from jets and satisfy $\pt>20$\GeV~(10\GeV) for the leading~(sub-leading) lepton and must be reconstructed within absolute pseudorapidity $|\eta|<2.5$.  
Exactly two same-flavor lepton candidates of opposite electric charge within a 10\GeV window around the \PZ boson mass, ${|m(\ell\bar\ell)-m_\PZ|<10\GeV}$, are required in each event.
The two leptons form the \PZ candidate.
Jets must satisfy $\pt>30$\GeV and $|\eta|<2.4$, and there must be either two or three jets, among which exactly two are required to be \cPqb~tagged.  
The \cPqb-tagged jets must satisfy $50<m(\textrm{\cPqb-jet}_1,\textrm{\cPqb-jet}_2)<150$\GeV and they form the \Ph candidate.
The Drell--Yan background process is also simulated in leptonic final states, including up to four extra jets. 
Potential SM-EFT effects of the Wilson coefficients on the background process are neglected.
For illustration, the reconstructed $p_{\textrm{T}}$ spectrum of the \PZ candidate is shown in~Fig.~\ref{fig:limits-delphes-binned}~(top left) where we have split the Drell--Yan process according to whether it contains a pair of \cPqb~partons within the kinematic acceptance at the ME level.
Despite the simplified \Delphes reconstruction, the cut-based analysis strategy, and differences in the object-level definitions, the signal-to-background ratios approximate the values from the nominal result of the CMS Collaboration closely enough that a proof-of-principle study is meaningful~\cite{CMS:2017odg}. 

An accurate estimate of the sensitivity requires, of course, the simulation of the relevant sub-leading backgrounds, a well-calibrated detector simulation, and a comprehensive assessment of systematic uncertainties in experimental, theoretical, and modeling aspects. 
These are beyond the scope of the present work.
Nevertheless, we assess the expected exclusion reach in this strongly simplified setting, despite these caveats.
We train the BIT estimators on half of the $6\times10^5$ simulated signal events and the $1.7\times 10^6$ background events with $B=200$, $N_\textrm{min}=50$, and $D=5$, as before. 
These events are not used for computing the exclusion reach in the following. 

{\renewcommand{\arraystretch}{1.3}
\begin{table}[t]
    \centering
    \caption{List of training variables in the \MGvATNLO study. We use the abbreviation $\Delta R=\sqrt{\Delta\eta^2+\Delta\phi^2}$ and $m$ denotes the invariant mass.} \label{tab:mva_observables}
    \begin{tabular}{c|c}
         Observable & Description \\\hline
         $H_{\textrm{T}}$&$\sum_{\textrm{jets}}\pt $\\
         $N_{\textrm{jet}}$&Jet multiplicity\\
         $\pt(j_1)$, $\pt(j_2)$, $\pt(j_3)$ & \pt of the  three highest \pt jets\\
         $|\eta(j_1)|$, $|\eta(j_2)|$, $|\eta(j_3)|$ & $|\eta|$ of the  three highest \pt jets \\
	$\pt(\Ph)$, $|\eta(\Ph)|$ & \pt and $|\eta|$ from \Ph candidate \\
	$\pt(\PZ)$, $|\eta(\PZ)|$ & \pt and $|\eta|$ from \PZ candidate \\
	$\Theta$, $\hat\theta$, $\hat\phi$ & See Sec.~\ref{sec:analytic-toy} and Fig.~\ref{fig:sketch}\\
$f_{LL}$, \ldots, $\widetilde f_{TT'}$ & See Sec.~\ref{sec:analytic-toy} and Ref.~\cite{Banerjee:2019twi}\\
    $\pt(\ell_2)/\pt(\ell_1)$ & Ratio of lepton \pt\\
	$\Delta\phi(\ell_1,\ell_2)$, $|\Delta\eta(\ell_1,\ell_2)|$  & Azimuthal and $\eta$ difference of $\ell_1$ and $\ell_2$ \\
	$\Delta\phi(\textrm{\cPqb-jet}_1,\textrm{\cPqb-jet}_2)$, $|\Delta\eta(\textrm{\cPqb-jet}_1,\textrm{\cPqb-jet}_2)|$  & Azimuthal and $\eta$ difference of \cPqb~jets\\
	$m(\textrm{\cPqb-jet}_1,\textrm{\cPqb-jet}_2)$& Higgs candidate mass\\
	$\pt(\textrm{\cPqb-jet}_2)/\pt(\textrm{\cPqb-jet}_1)$ & Ratio of transverse \cPqb-jet momenta\\
   $\Delta R(\PZ,\Ph)$, $|\Delta \eta(\PZ,\Ph)|$, $m(\PZ, \Ph)$ & Properties of the $\PZ\Ph$ system\\
   $\Delta R(\textrm{non \cPqb-jet}, \PZ)$,  $\Delta R(\textrm{non \cPqb-jet}, \Ph)$ & $\Delta R$ distances to non \cPqb-tagged jet\\
      Thrust & See Ref.~\cite{PhysRevLett.39.1587}\\
    \end{tabular}
\end{table}
}

We already know from Sec.~\ref{sec:toys-exclusion-binning} that $p_{\textrm{T,Z}}$ and the angular observables in Fig.~\ref{fig:sketch} can capture the relevant SM-EFT effects in the absence of backgrounds. 
When training with combined background and signal samples, the background contributions and the SM-EFT sensitivity are simultaneously encoded in the prediction of $R(\boldsymbol{x}|\boldsymbol{\theta},\boldsymbol{\theta}_0)$.
To facilitate the SM-EFT exclusion power, observables sensitive to differences between the Drell--Yan and $\PZ\Ph$ processes are included. 
These comprise the pseudorapidity of the jets, their transverse momenta and scalar sum~($H_{\textrm{T}}$), the transverse momentum and ratios of the leptons and \cPqb-tagged jets, and angular observables as well as the invariant mass computed from reconstructed $\PZ$ and $\Ph$ momenta. 
The analysis selection requires exactly two \cPqb~jets that form the \Ph candidate and, therefore, the \cPqb-jet multiplicity is not in the list of training variables, summarized in Table~\ref{tab:mva_observables}.

Figure~\ref{fig:limits-delphes-binned} shows the power of the resulting BIT estimate $\hat q_{\boldsymbol{\theta}}$ in the  $C_{\PH\PQ^{(3)}}$--$C_{\PH\PW}$ plane~(top right), the $C_{\PH\PW}$--$C_{\PH\widetilde{\PW}}$ plane~(bottom left), and the  $C_{\PH\PQ^{(3)}}$--$C_{\PH\widetilde{\PW}}$ plane~(bottom right). 
The black contour shows the median expected exclusion reach at 95\% CL for an $M=30$ binning, constructed as described in Sec.~\ref{sec:toys-exclusion-binning}. 
The presence of the backgrounds  reduces the expected sensitivity significantly.
The shapes of the contours are similar to the toy study in Sec.~\ref{sec:toys-exclusion-binning}.
The binned version can also be used to get some insight into the relative importance of the linear and the quadratic terms by dropping the other term in $\hat R$ in Eq.~\ref{eq:F-poly-estimation}.
While non-zero linear coefficients will always lead to regions in parameter space where the linearly truncated $\hat R(\boldsymbol{x}|\boldsymbol{\theta},\boldsymbol{\theta}_0)$ takes unphysical negative values, these do not stop us from constructing the binning and, subsequently, evaluating Eq.~\ref{eq:binned-test-stat}. 
Removing the linear term, on the other hand, can not change the sign of $R(\boldsymbol{x}|\boldsymbol{\theta},\boldsymbol{\theta}_0)$.
In Fig.~\ref{fig:limits-delphes-binned}, we also show the 95\%~CL exclusion contours as obtained from the binned test statistic using the linear~(blue lines) and quadratic~(red lines) terms.
The relative importance of the terms depends on the parameter point, but the all-order nominal BIT estimate, leads to better performance in all cases. 
If the validity of the SM-EFT expansion beyond the linear term is a concern, the binned version of the linearized $\hat R(\boldsymbol{x}|\boldsymbol{\theta},\boldsymbol{\theta}_0)$ offers an estimator whose training does not depend on the coefficients in doubt, and is still optimal, provided $\boldsymbol{\theta}$ and $\boldsymbol{\theta}_0$ are close. 
The choice can be made after the training.

\begin{figure}\centering
        \hfill        
        \includegraphics[width=0.49\textwidth]{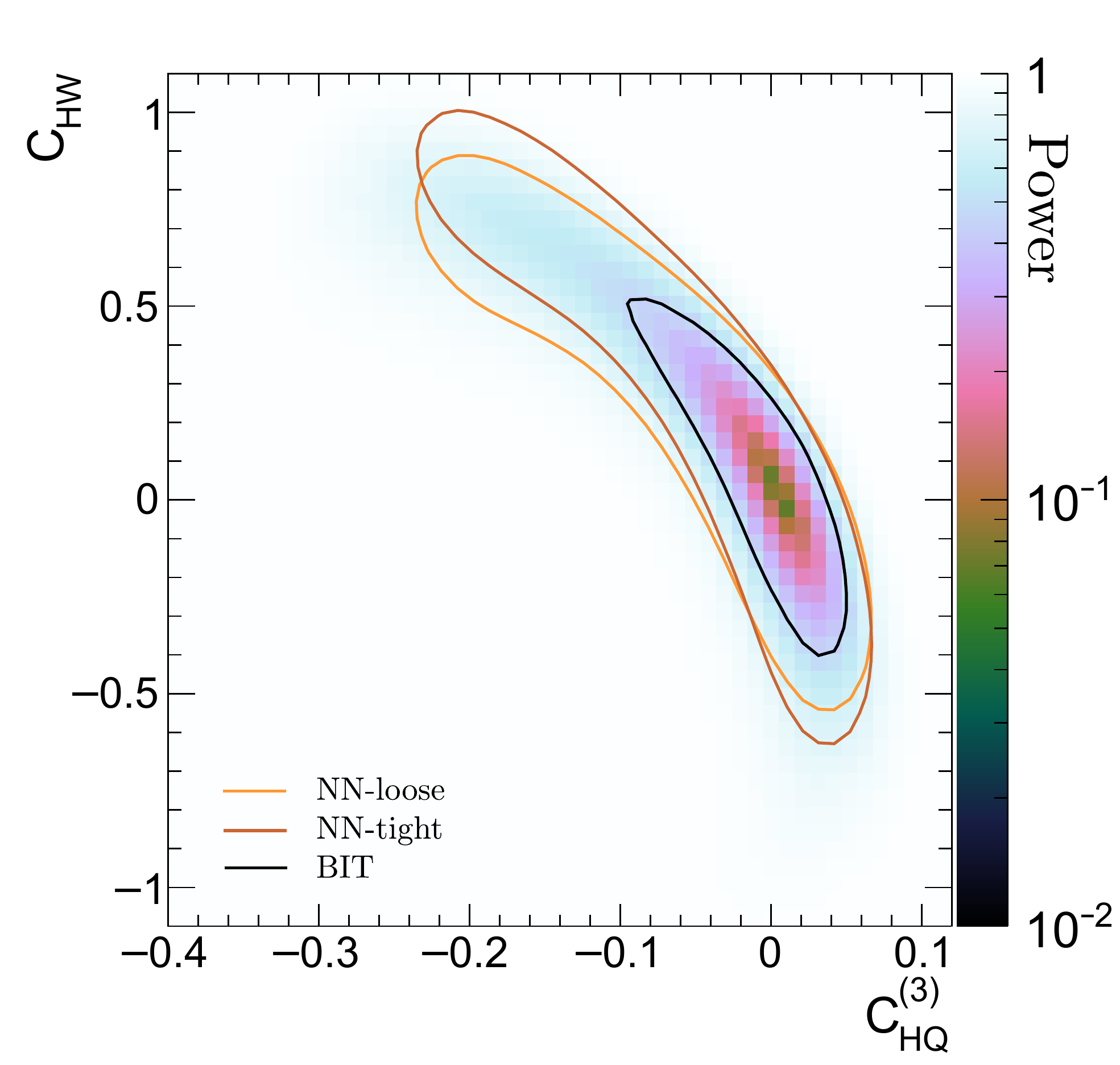}\\
        \raisebox{0mm}{\includegraphics[width=0.49\textwidth]{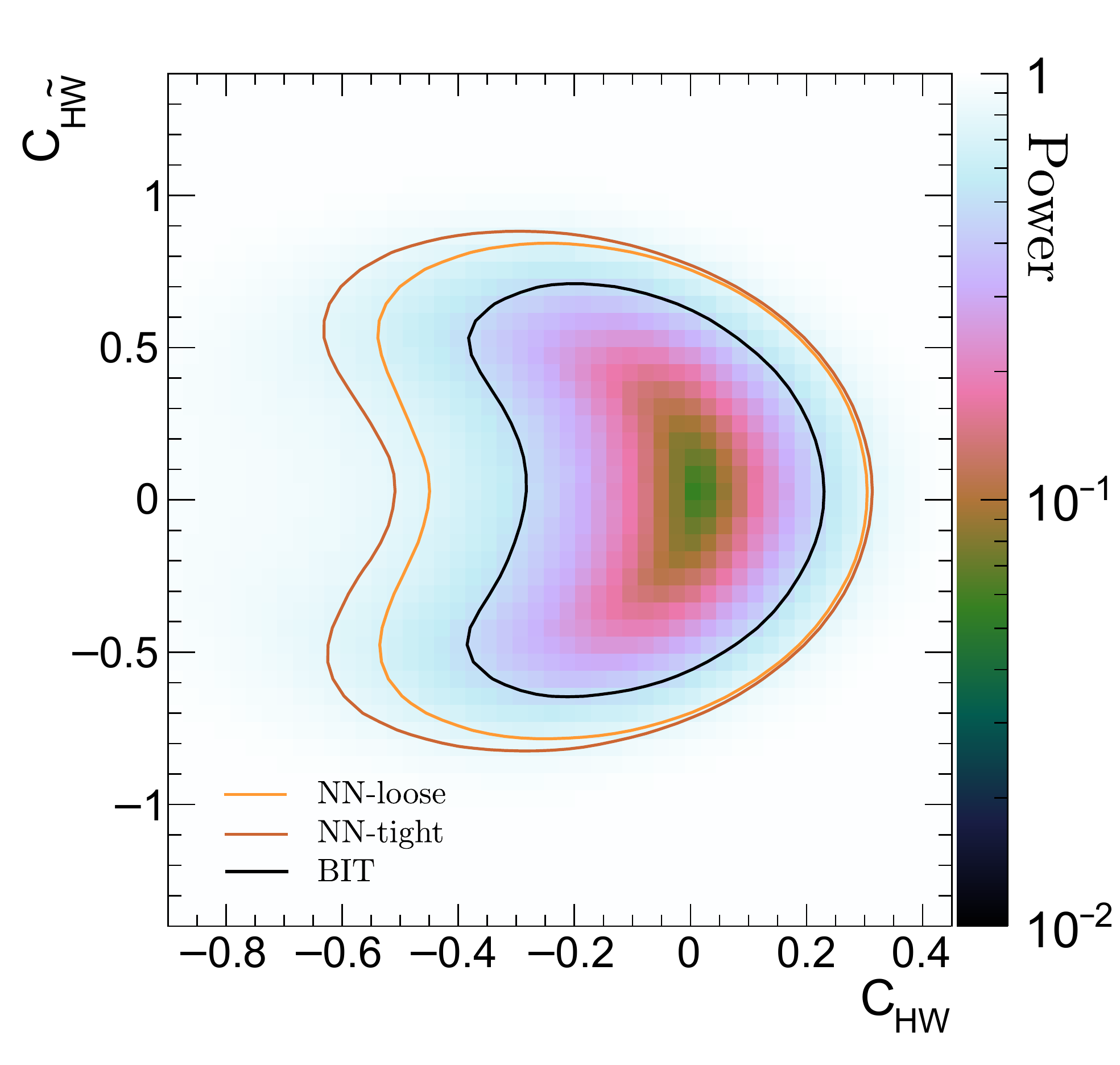}}
        \includegraphics[width=0.49\textwidth]{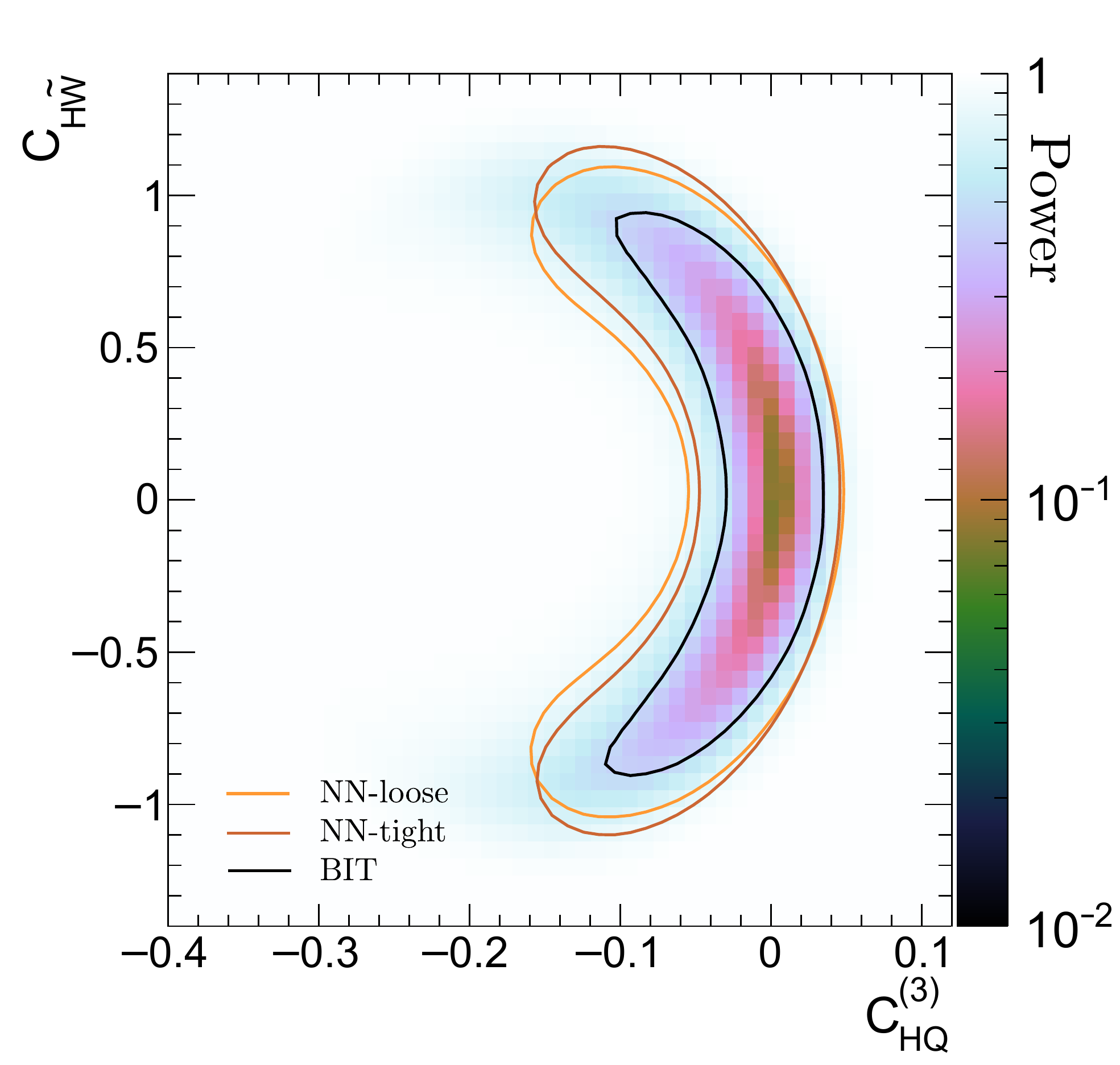}
    \caption{
    Power of the $\hat q$ test statistic as in Fig.~\ref{fig:limits-delphes-binned} and the median expected exclusion countour at 95\%CL~(black). The colored contours show the median expected exclusion limit for the $p_{\textrm{T,Z}}$ test statistic after a loose or a tight requirement on the classification NN described in the text.}\label{fig:limits-ATS-binned}
\end{figure}
Motivated by recent analyses~\cite{ATLAS:2020fcp,ATLAS:2020jwz} in the STXS binning scheme~\cite{deFlorian:2016spz,Andersen:2016qtm,Berger:2019wnu} that effectively adopt a coarse $p_{\textrm{T,Z}}$ binning, we finally compare the BIT approach to a simple $p_{\textrm{T,Z}}$-based strategy.
For background suppression, we use the \textsc{Keras} package~\cite{chollet2015keras} to train a neural network~(NN) classifier separating the Drell--Yan background and the $\PZ\Ph$ signal.
We use the same input features as for the BIT training, and configure the network with two fully connected internal layers, consisting of 100 and 50 neurons, respectively.
The neurons' activation functions are chosen as sigmoid, and we adopt the \textsc{Adam} optimizer with the mean-squared error loss function.
We have checked that different network configurations and alternative loss functions do not significantly change the performance, provided the number of neurons is not too small. 
Given the shortcomings of the simulation, a more detailed optimization is not our priority. 
The resulting classifier is used to define a loose~(tight) working point with a signal efficiency of 77\%~(33\%) and a background efficiency of 25\%~(3.5\%). 
With the corresponding threshold requirement on the NN classifier in place, we use $p_{\textrm{T,Z}}$ instead of $\hat R(\boldsymbol{x}|\boldsymbol{\theta},\boldsymbol{\theta}_0)$ to define an $M=30$ binning. The resulting contours are shown in Fig.~\ref{fig:limits-ATS-binned}. We have checked that intermediate NN working points lead to exclusion contours between the contours of the loose and the tight NN working point. 
The BIT approach leads to significantly better reach in this example, which we attribute to the fact that it can exploit the SM-EFT dependence of the differential cross section ratio in the full feature space.

\section{Conclusion}\label{sec:conclusion}

In this work, we have developed a tree-boosting algorithm for multi-parameter measurements of the Wilson coefficients in the context of the SM-EFT and, potentially, other effective field theories.
The algorithm trains regressors in the coefficient functions of the differential cross section ratio in the expansion in the Wilson coefficients. 
If this expansion terminates, which is usually the case if the SM-EFT dependence is obtained at fixed order in perturbation theory, a small number of such regressors is sufficient to provide an optimal test statistic in the sense of the Neyman-Pearson lemma. 
The resulting test statistic has simple analytic dependence on the coefficient functions and can be used in both binned or unbinned hypothesis tests.
The approach allows including SM-EFT effects order-by-order, and the number of required coefficient functions corresponds to the number of degrees of freedom in the SM-EFT expansion.

Technically, we build on our earlier work in Ref.~\cite{Chatterjee:2021nms}. We use the weak learner of a CART algorithm for estimating the higher-order terms in the $\boldsymbol{\theta}$-expansion of the cross section ratio. 
Using well-known MSE loss-functions, we derive the boosting algorithm that leverages weight functions of simulated event samples, encoding the SM-EFT effects in the whole parameter space.

With toy simulation of an analytic model of the $\PZ\Ph$ process, we assess the algorithm's optimality and compare binned and unbinned exclusion reaches.
Furthermore, we apply the algorithm to simulation using a Monte-Carlo event generator and parameterized detector reconstruction, including the most important backgrounds. 
We use this simplified simulation to show that the our approach provides significant improvement in SM-EFT exclusion reach, when compared to a simple $p_{\textrm{T,Z}}$-based strategy that employs a neural network trained with the same information to achieve background suppression. 
\clearpage
\section*{Acknowledgements}
We are grateful to Shankha Banerjee and Rick S. Gupta for providing the translation of the SM-EFT parametrization into the conventions of Ref.~\cite{Nakamura:2017ihk} and the formulae for the contact interactions. 
\textit{Funding:} The work of D. S. was supported by the Austrian Science
Fund (FWF) project P33771.

\bibliographystyle{elsarticle-num}
\bibliography{references}

\appendix
\clearpage
\section{The weak learner}\label{app:weak-lerner}
In Algorithm~\ref{alg:weak-learner}, we present the pseudo-code for the fitting procedure of our decision-tree weak learner.
It is the same as Algorithm~1 in our earlier work Ref.~\cite{Chatterjee:2021nms} except for the input weight coefficients that can now either be $\{w_{a,i}\}_{i=1}^N$ or $\{w_{ab,i}\}_{i=1}^N$. 
The algorithm is an adaption of the ``Classification And Regression Tree''~(CART) algorithm by Breiman et al.~\cite{breiman1984classification}, where the feature space in each recursion step is further divided by greedily selecting the dimension and cut value combination that maximizes a local gain. 
It can be efficiently implemented by sorting the events' features for each dimension separately and using cumulative sums over the weights and weight derivatives to find the locally optimal cut. 
The terminal nodes $j \in \mathcal J$ are encoded each as tuple consisting of the requirements $\alpha_j$ (selecting a subset of the data $\mathcal{D}_{\alpha} \subset \mathcal{D}$) and prediction $F_j$.
They determine the final weak learner $F(\mathbf{x}) = \sum_{j \in \mathcal{J}} \mathds{1}_{\alpha_j}(\mathbf{x}) \cdot F_j$.
Overfitting and computational complexity are discussed in Ref.~\cite{Chatterjee:2021nms}. 
\begin{algorithm}[h]
    \setstretch{1.1}
	\KwData{Data set $\mathcal{D} = \{\mathbf{x}_i,w_{i,0},w_{a,i}\}_{i=1}^N$ for fixed $a$ or $\mathcal{D} = \{\mathbf{x}_i,w_{i,0},w_{ab,i}\}_{i=1}^N$ for fixed $ab$.}
	\KwIn{Tree depth $D$, minimum terminal node size $N_\mathrm{min}$}
	\KwOut{Weak learner $F$ \vspace{0.02cm}}

    $w'\gets \{w_{a,i}\}_{i=1}^N$ or $\{w_{ab,i}\}_{i=1}^N$ \{drop $a$ and $ab$ to reduce clutter\}
    
	$\alpha \gets (), Q \gets (\alpha), \mathcal{J} \gets \{\}, P \gets \{1, \dots, d\}$\;
	
	\While{$Q \ne \emptyset$}{
		$\alpha \gets \mathsf{pop}(Q)$\;
		
		$\bm \pi \gets \mathrm{arg\,sort}\ \mathbf{x} \in \mathcal{D}_{\alpha}$ \algorithmiccomment{sorted indices for each dimension $p \in P$}\;
		
		\uIf{$|\alpha| \le D \land |\mathcal{D}_{\alpha}| > 2N_{\textrm{min}}$}{
		    $K \gets \{N_{\textrm{min}},\dots, |\mathcal{D}_{\alpha}| - N_{\textrm{min}}\}$\; \algorithmiccomment{allowed cut points}
		    
			$p^*, k^* \gets \argmax_{p\in P, k\in K} \left[\frac{\left(\sum_{i=1}^k w'_{\bm \pi_{p, k}}\right)^2}{ \sum_{i=1}^k w_{\bm \pi_{p, k}}} + \frac{\left(\sum_{i=k+1}^{|\mathcal{D}_{\alpha}|} w_{\bm \pi_{p, k}}\right)^2}{\sum_{i=k+1}^{|\mathcal{D}_{\alpha}|} w_{\bm \pi_{p, k}}}\right]$\;
			
			$c \gets x_{\bm \pi_{p^*,k^*},p^*}$ from $\mathcal{D}_{\alpha}$\;
			
			\uIf{$c$ is a valid cut}{
				$\alpha_\mathrm{L} \gets \alpha \cup (p^*, \le, c)$,\;	$\alpha_\mathrm{R} \gets \alpha \cup (p^*, >, c)$,\;
				$Q \gets Q \cup (\alpha_\mathrm{L}, \alpha_\mathrm{R})$\;
			}
			\Else{
				$\mathcal{J} \gets \mathcal{J} \cup \left(\alpha, \frac{\sum_{w' \in \mathcal{D}_{\alpha}} w'}{ \sum_{w \in \mathcal{D}_{\alpha}} w }\right)$\;
			}	
		}
		\Else{
			$\mathcal{J} \gets \mathcal{J} \cup \left(\alpha, \frac{\sum_{w' \in \mathcal{D}_{\alpha}} w' }{\sum_{w \in \mathcal{D}_{\alpha}} w }\right)$\;
		}		
	}
	\Return $F(\mathbf{x}) = \sum_{j \in \mathcal{J}} \mathds{1}_{\alpha_j}(\mathbf{x}) \cdot F_j$\;
	
	\caption{The $\mathsf{fit}$ procedure for the weak learner } 
	\label{alg:weak-learner}
\end{algorithm}


\end{document}